\documentclass[a4paper,twocolumn,11pt]{quantumarticle}
\pdfoutput=1
\usepackage[utf8]{inputenc}
\usepackage[english]{babel}
\usepackage[T1]{fontenc}
\usepackage{amsmath,amssymb,amsfonts}
\usepackage{mathtools}
\usepackage{amsthm}             

\usepackage{braket}                    
     
\usepackage{etoolbox}
\usepackage{hyperref}
\usepackage{float}
\usepackage{booktabs}
\usepackage{xspace}  
\usepackage{enumitem}
\usepackage{multirow}

\DeclareRobustCommand{\SC}[1]{\textsc{\MakeUppercase{#1}}}

\newcommand{\Nstep}[1]{#1-step\xspace}

\newcommand{\sixstep}{\Nstep{six}}

\newcommand{\FCfull}{Floquet color}

\newcommand{\HF}{\SC{HF}\xspace}

\newcommand{\HCF}{\SC{HCF}\xspace}
\newcommand{\HCFfull}{hyperbolic color Floquet}

\newcommand{\plaquette}{plaquette\xspace}

\newcommand{\MWPM}{\SC{MWPM}\xspace}

\newcommand{\BPOSD}{\SC{BP+OSD}\xspace}

\newcommand{\EMthree}{\SC{EM3}\xspace}

\newcommand{\EMind}{\EMthree -ind\xspace}
\newcommand{\EMcor}{\EMthree -cor\xspace}

\newcommand{\XZ}[2]{\ensuremath{X^{#1}\,Z^{#2}}}
\newcommand{\XYZ}[3]{\ensuremath{X^{#1}\,Y^{#2}\,Z^{#3}}}

\DeclareRobustCommand{\XoX}{\ensuremath{X\!\otimes\!X}}
\DeclareRobustCommand{\YoY}{\ensuremath{Y\!\otimes\!Y}}
\DeclareRobustCommand{\ZoZ}{\ensuremath{Z\!\otimes\!Z}}

\newcommand{\graphedge}{graph-edge\xspace}
\newcommand{\Graphedge}{Graph-edge\xspace}
\newcommand{\graphedgesyn}{graph-edge syndrome\xspace}
\newcommand{\graphedgesyns}{graph-edge syndromes\xspace}

\newcommand{\Hyperedge}{Hyperedge\xspace}
\newcommand{\hyperedgesyn}{hyperedge syndrome\xspace}

\NewDocumentCommand{\dist}{}{%
  \ensuremath{d}\xspace
}
\NewDocumentCommand{\PL}{}{\ensuremath{P_\mathrm{L}}\xspace}
\NewDocumentCommand{\Pphys}{}{\ensuremath{P_\mathrm{phys}}\xspace}

\NewDocumentCommand{\codedistance}{s}{%
  code distance%
  \IfBooleanF{#1}{\ \dist}%
}

\NewDocumentCommand{\logicalrate}{s}{%
  \IfBooleanTF{#1}{logical error rate\xspace}{logical error rate \PL}%
}

\NewDocumentCommand{\Logicalrate}{s}{%
  \IfBooleanTF{#1}{Logical error rate\xspace}{Logical error rate \PL}%
}

\NewDocumentCommand{\physicalrate}{s}{%
  \IfBooleanTF{#1}{physical error rate\xspace}{physical error rate \Pphys}%
}

\newcommand{\finiteenc}{finite encoding rate\xspace}

\newcommand{\Encodingrate}{Encoding rate\xspace}

\newcommand{\wt}[1]{weight-#1\xspace} 
\newcommand{\body}[1]{#1-body\xspace}

\newcommand{\weighttwo}{\wt{2}}
\newcommand{\twobody}{\body{2}}

\newcommand{\paritychecks}{parity checks\xspace}

\newcommand{\semihyperbolic}{semi-hyperbolic\xspace}

\newcommand{\threecolorable}{three-colorable\xspace}

\newcommand{\qubit}{qubit\xspace}
\newcommand{\qubits}{qubits\xspace}

\newcommand{\threeD}{\mbox{3D}\xspace}

\newcommand{\singlefaultevent}{single-fault event\xspace}

\newcommand{\vs}{vs.\@\xspace}

\newcommand{\eg}{e.g.,\@\xspace}

\newcommand{\ie}{i.e.,\@\xspace}

\begin{document}

\title{Hyperbolic Floquet code with graph-edge syndromes}

\author{Hideyuki Ozawa}
\affiliation{Information Technology R\&D Center, Mitsubishi Electric Corporation, Kamakura 247-8501, Japan}

\author{Isamu Kudo}
\affiliation{Information Technology R\&D Center, Mitsubishi Electric Corporation, Kamakura 247-8501, Japan}

\author{Yuki Takeuchi}
\affiliation{Information Technology R\&D Center, Mitsubishi Electric Corporation, Kamakura 247-8501, Japan}

\author{Tsuyoshi Yoshida}
\affiliation{Information Technology R\&D Center, Mitsubishi Electric Corporation, Kamakura 247-8501, Japan}

\maketitle

\begin{abstract}

Quantum error correction would be a primitive for demonstrating quantum advantage in a realistic noisy environment. Floquet codes are a class of dynamically generated, stabilizer-based codes in which low-weight parity measurements are applied in a time-periodic schedule. Furthermore, for several Floquet codes, the encoding rate becomes finite even for an infinitely large \qubit number. However, despite these advantageous properties, existing Floquet codes require handling more intricate, often hypergraph-structured syndromes from the decoding perspective, which makes decoding comparatively demanding. We give a concrete method for solving this issue by proposing hyperbolic color Floquet (\HCF) code. To this end, we simultaneously take advantage of hyperbolic Floquet and Floquet color codes. Parity measurements in our code consist of the repetitions of six-step measurements on (semi-)hyperbolic three-colorable tilings. Since each step just measures $X \otimes X$ or $Z \otimes Z$, our code on the regular $\{8,3\}$ lattice has the following three advantages: (i) parity measurements are weight-2, (ii) for the numbers $k$ and $n$ of logical and physical \qubits, respectively, the encoding rate is finite, \ie $\lim_{n \to \infty} k/n = 1/8$, and (iii) the code distance is proportional to $\log n$. From the above property, each \singlefaultevent\ generally affects at most two detectors, which implies ``graph-edge'' syndromes, and hence decoding with a minimum-weight perfect matching (\MWPM) decoder is efficient and virtually scales near-linearly in the number of physical \qubits $n$. This is a stark contrast to several known Floquet codes because their parity measurements repeat the measurements of \XoX, \YoY, and \ZoZ, and thus the syndromes are represented as a hypergraph, which basically requires decoders with longer decoding time. It is worth mentioning that our HCF code can be straightforwardly defined on any trivalent lattice in hyperbolic geometry.

\end{abstract}

%======================================================================
\section{Introduction}
\label{sec:new_intro}
%======================================================================
There exist quantum algorithms, executable on a universal quantum computer, that solve certain problems asymptotically faster than the best-known classical algorithms.
These algorithms include prime factorization~\cite{Shor1997}, simulations of chemical and condensed-matter systems~\cite{Georgescu2014,Gharibian2022}, and certain machine-learning tasks~\cite{Harrow2009}. Motivated by these computational advantages, tremendous efforts have been devoted to realizing large-scale universal quantum computers. However, \qubits are fragile and can easily lose their quantum properties due to decoherence and control errors. Therefore, countermeasures against noise are essential primitives for realizing quantum computers.

Two promising approaches are quantum error mitigation (QEM)~\cite{Cai2023} and quantum error correction (QEC)~\cite{Nielsen2010}. The former reduces errors through postprocessing or by repeating quantum operations without encoding into logical qubits, making it compatible with currently available noisy intermediate-scale quantum (NISQ) devices~\cite{Preskill2018}. However, QEM is generally not expected to be scalable; for example, recent analyses indicate that even under depolarizing noise models, the sampling overhead required to reach a target accuracy can grow exponentially with circuit depth~\cite{Takagi2022,Tsubouchi2023,Takagi2023,Quek2024}. In contrast, QEC is in principle scalable with respect to algorithm size and depth, though it requires encoding into logical qubits; thus, it is compatible with large-scale universal quantum computers. Hybrid approaches that combine QEC with QEM have also been explored as a means to alleviate the practical challenges of QEC~\cite{Suzuki2022}.

Over the past quarter-century, QEC has progressed from a purely theoretical 
construct to an experimentally demonstrated component of quantum information 
processing. On the theoretical side, the stabilizer formalism and Calderbank-Shor-Steane (CSS) construction~\cite{Gottesman1997Stabilizer,Knill1997Theory,Calderbank1996Good} laid the foundation for the field. These were followed by topological surface and color codes~\cite{Dennis2002Topological,Bombin2006}, which offer 
high-threshold, geometrically local protection at the cost of a vanishing 
encoding rate. More recent advances in quantum low-density parity-check (LDPC) codes~\cite{Panteleev2021Goodcode} achieve both finite encoding rates and linear distances, although their high-weight parity checks remain challenging for near-term hardware.

Subsystem (gauge) codes and gauge fixing 
techniques~\cite{Bacon2006Operator,Paetznick2013Universal,Chamberland2020Topological} 
reduce measurement weight and enable transversal logical gates, while bosonic and concatenated architectures further expand the range of available QEC approaches.
The construction of novel QEC codes is important not only for realizing reliable large-scale quantum computers but also for advancing fundamental theory. For instance, the no low-energy trivial state (NLTS) conjecture—an ingredient toward the quantum probabilistically checkable proof (PCP) conjecture~\cite{Aharonov2013}—was recently resolved using quantum LDPC codes~\cite{Anshu2023}.

Experimentally, successive prototypes have observed near-exponential suppression of logical error rates 
with increasing code distance in superconducting 
\qubits~\cite{Krinner2022ErrorBudget,Arute2019Supremacy} and trapped ions~\cite{Wright2019Benchmarking,RyanAnderson2021RealTime,Egan2021FaultTolerant}.

Floquet codes employ a time-periodic gauge fixing schedule that periodically measures 
\weighttwo\ \paritychecks, thereby generating a dynamic stabilizer (or gauge) 
group~\cite{HastingsHaah2021Dynamical}. In prior Floquet code studies, geometry 
(Euclidean vs.\ hyperbolic) and measurement basis (the three-Pauli basis 
\(\{\XoX,\YoY,\ZoZ\}\) vs.\ the two-Pauli basis \(\{\XoX,\ZoZ\}\)) have often 
been treated independently. 
Euclidean constructions such as the honeycomb code~\cite{Vuillot2021Planar,Gidney2022Planar,Haah2022Boundaries} and \FCfull\ code~\cite{Davydova2023,Kesselring2024} preserve 
weight-2 interactions. However, the number of logical \qubits\ remains constant 
(e.g., \(k=1\)), causing the encoding rate to vanish asymptotically ($k/n \to 0$). 
Hyperbolic constructions—typified by the hyperbolic Floquet (\HF) code~\cite{Higgott2024SemiHyper,Fahimniya2024} and the semi-hyperbolic Floquet code 
—remain \finiteenc\ (\(k/n=\Theta(1)\)) but at the cost of introducing 
a three-Pauli basis that increases decoding complexity.

To address these limitations, we propose \HCFfull\ (\HCF) code, which 
adapts the two-Pauli basis $\{X \otimes X, Z \otimes Z\}$ to hyperbolic, 
three-colorable tilings—specifically, $\{8,3\}$ tilings—and their 
semi-hyperbolic refinements, achieving finite encoding rates with $k/n=\Theta(1)$. 
Under circuit-level depolarizing noise, the HCF schedule typically constrains single-fault events to trigger at most two detection events, enabling minimum-weight perfect matching (MWPM) to achieve near-linear decoding time in $n$, empirically $\approx O(n)$~\cite{Higgott2025sparseblossom}. Compared with BP-based decoders, this near-linear scaling yields a lower and more practically useful computational cost.

\bigskip
The organization of this paper is as follows:

\begin{itemize}[leftmargin=1.6em]
  \item Section~\ref{sec:hyper_geom} introduces hyperbolic geometry, 
        regular $\{p,3\}$ tilings, and the semi-hyperbolic refinement 
        that enhances distance scaling.
  \item Section~\ref{sec:landscape} provides an overview of existing 
        Floquet codes and positions \HCF\ code within that landscape.
  \item Section~\ref{sec:anyon_overview} presents an anyon-theoretic 
        description of Floquet dynamics and analyzes the condensations 
        relevant to \HCF\ code.
  \item Section~\ref{sec:code_construction} details the construction 
        of \HCF\ code, including its six-step two-body measurement 
        schedule, detector lattice, and logical-operator dynamics.
  \item Section~\ref{sec:noise_decoder} describes the noise models and 
        candidate decoders, and introduces the algorithms employed 
        in our work.
  \item Section~\ref{sec:calculation} presents numerical calculations 
        including finite-size scaling, decoder dependence, detector weights, and threshold estimates analysis.
  \item Section~\ref{sec:summary} provides conclusions and directions for future work.
\end{itemize}

%======================================================================
\section{Hyperbolic geometry and tilings}\label{sec:hyper_geom}
%======================================================================
This section presents the geometric background required for \HCF\ code construction: regular hyperbolic tilings and the semi-hyperbolic refinement.

%-----------------------------------------------------------------------
\subsection{Hyperbolic preliminaries}\label{subsec:hyper_prelims}
\begin{figure}[H]
\centering
\includegraphics[width=\linewidth]{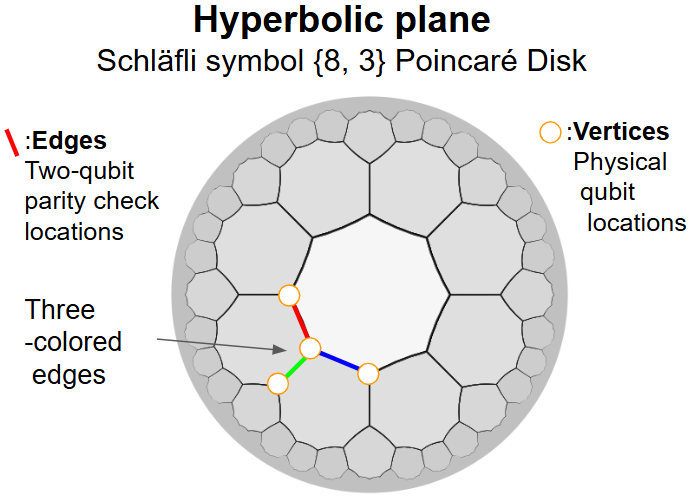}
\caption{Poincaré projection of the $\{8,3\}$ tiling by regular octagons. 
       Orange circles mark physical qubit positions at the vertices. 
       Black edges indicate the locations of two-body parity checks. 
       The three-colored edges enable a periodic measurement schedule.}
\label{fig:hyperbolic_lattice}
\end{figure}

%-----------------------------------------------------------------------
Throughout, we adopt the Poincaré disk model
\begin{equation}
  \mathbb{D}=\bigl\{z\in\mathbb{C}\,\bigm|\,|z|<1\bigr\},
\end{equation}
whose geodesics are circular arcs orthogonal to the boundary
$\partial\mathbb{D}$. A tiling is called regular when each face
is a regular $p$-gon and $q$ faces meet at every vertex. 
Such tilings are denoted $\{p,q\}$ (see Figure~\ref{fig:hyperbolic_lattice}). 
Negative curvature arises whenever
\begin{equation}
  \frac{1}{p}+\frac{1}{q}<\frac{1}{2},
\end{equation}
implying exponential area growth with graph radius—a key ingredient 
for the \finiteenc behavior of hyperbolic Floquet codes.

%-----------------------------------------------------------------------
\subsection{Regular \texorpdfstring{$\{p,q\}$}{\{p,q\}} tilings}
\label{subsec:regular_pq}
%-----------------------------------------------------------------------
Let $(V,E,F)$ denote the vertex, edge, and face sets, respectively, of a compact 
$\{p,q\}$ tiling. Double-counting edge-vertex and edge-face incidences yields
\begin{equation}
  q|V| = 2|E|, \qquad p|F| = 2|E|.
  \label{eq:regular_count}
\end{equation}
Placing the tiling on a closed surface of genus $\mathfrak{g}$ and 
applying Euler's formula $|V| - |E| + |F| = 2 - 2\mathfrak{g}$ yields
\begin{equation}
  \mathfrak{g} = 1 - |E|\left(\frac{1}{p} + \frac{1}{q} - \frac{1}{2}\right).
  \label{eq:regular_genus}
\end{equation}

%-----------------------------------------------------------------------
\subsection{Three-color tiling}
\label{subsec:three_colorability}
%-----------------------------------------------------------------------
A tiling $T=(V,E,F)$ is called a three-color tiling if it has 
trivalent vertices and its faces admit a three-coloring where adjacent faces 
have distinct colors. Throughout this work, we focus on the hyperbolic 
tiling with Schläfli symbol $\{8,3\}$, which has vertex configuration 
$(8.8.8)$ (see Figure~\ref{fig:hyperbolic_lattice}). 
Note that the hyperbolic tilings with vertex configurations $(4.8.10)$ and $(4.10.10)$ also 
satisfy these conditions~\cite{Higgott2024SemiHyper}.

\begin{figure}[H]
  \includegraphics[width=\linewidth]{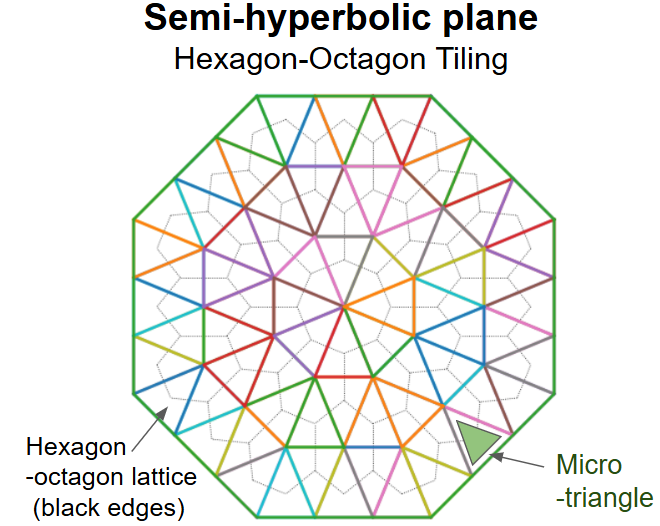}
  \caption{Refinement of the $\{8,3\}$ octagon lattice with $\ell=3$. 
    The Poincaré disk illustrates three successive transformations: 
    (1) octagon-triangle decomposition, where each octagon is cut into eight triangles; 
    (2) $\ell$-fold subdivision, where each triangle is replaced with $\ell^2$ 
    micro-triangles (one example highlighted with thick colored edges); 
    (3) hexagon-octagon \emph{dual} rebuild, where the refined triangulation is 
    converted into a mixed hexagon-octagon lattice (black edges).}
  \label{fig:semi_hyp_schema}
\end{figure}

%-----------------------------------------------------------------------
\subsection{Semi-hyperbolic refinement}
\label{subsec:semi_hyperbolic}
%-----------------------------------------------------------------------
To enhance the logarithmic distance scaling of hyperbolic codes, 
the semi-hyperbolic refinement procedure can be employed as follows:
(i)~each $p$-gon is decomposed into eight triangles;
(ii)~an $\ell \times \ell$ subdivision is performed for $\ell \geq 1$;
(iii)~a dual transformation is applied to obtain a trivalent lattice 
(see Figure~\ref{fig:semi_hyp_schema}).
This procedure preserves the trivalent structure while yielding the scaling relations
\begin{equation}
  n_\ell = \Theta(\ell^2 n), \qquad d_\ell = \Theta(\ell d),
  \label{eq:sh_scaling}
\end{equation}
where $n_\ell$ and $d_\ell$ are the number of qubits and code distance 
after refinement with subdivision parameter $\ell$, respectively, 
while $n$ and $d$ denote the corresponding initial values.

% ===============================================================
\section{Floquet code landscape}\label{sec:landscape}
% ===============================================================
%-----------------------------------------------------------------------
\subsection{Basics of quantum error correction}
%-----------------------------------------------------------------------

\paragraph*{Stabilizer and CSS codes}
Stabilizer codes~\cite{Gottesman1997Stabilizer,Gottesman1998Theory} are 
quantum error-correcting codes whose logical subspace is the joint $+1$ 
eigenspace of commuting Pauli operators that form the stabilizer group. 
The formalism applies to block encodings over both qubits and qudits. 
It provides standardized approaches for encoding, gate operations, and 
syndrome-based decoding through stabilizer measurements and Pauli error 
correction.

CSS codes~\cite{Calderbank1996Good,Steane1996Error} feature stabilizer 
generators that separate into $X$-type and $Z$-type operators. 
These two sets often correspond to components of a chain complex over an appropriate ring or field.

\paragraph*{Subsystem (gauge) codes}
Subsystem (gauge) codes~\cite{Bacon2006Operator,Poulin2005Stabilizer,Bombin2015Gauge} 
are specified by a gauge group $\mathsf{G}\subset\mathsf{P}_n$, where 
$\mathsf{P}_n$ denotes the $n$-qubit Pauli group. The stabilizer subgroup 
consists of the central elements of the gauge group:
\begin{equation}
  \mathsf{S}=\mathsf{Z}(\mathsf{G})\cap\mathsf{G},
  \label{eq:G1}
\end{equation}
where $\mathsf{Z}(\mathsf{G})$ denotes the center of $\mathsf{G}$. 
The logical Paulis are given by the quotient
\begin{equation}
  \mathsf{L}=\mathsf{N}(\mathsf{S})/\mathsf{G},
  \label{eq:G2}
\end{equation}
where $\mathsf{N}(\mathsf{S})=\{P\in\mathsf{P}_n : PS=SP \text{ for all } S\in\mathsf{S}\}$ 
is the normalizer of $\mathsf{S}$ in $\mathsf{P}_n$. The code distance is defined as the minimum weight of a non-trivial logical operator:
\begin{equation}
  d=\min\bigl\{\mathrm{wt}(P) : P\in \mathsf{N}(\mathsf{S})\setminus\mathsf{G}\bigr\},
  \label{eq:G3}
\end{equation}
where $\mathrm{wt}(\cdot)$ denotes the Pauli weight. Here, "$\setminus \mathsf{G}$" denotes the set difference, excluding gauge operators such that only logical operators that act non-trivially on the logical subsystem are considered. 
Notable examples include operator versions of Bacon-Shor codes~\cite{Bacon2006Operator}, 
gauge color codes with gauge fixing~\cite{Bombin2015Gauge}, and fault-tolerant constructions with low-degree topological 
realizations~\cite{Chamberland2020Topological,Brown2016Fault}.

\paragraph*{Floquet and dynamical codes}
Dynamical codes~\cite{HastingsHaah2021Dynamical,FuGottesman2024ECinDynamicalCodes} are defined by a sequence of Pauli measurements, called a measurement schedule. 
Let $\{\mathcal{M}_t\}_{t\geq 1}$ denote the schedule, with the instantaneous stabilizer group (ISG) $\mathsf{S}_t$ determined from measurements up to time $t$. 
The corresponding code space and logical operators are
\begin{equation}
  \mathcal{Q}_t=\mathcal{Q}(\mathsf{S}_t), \qquad
  \mathsf{L}_t=\mathsf{N}(\mathsf{S}_t)/\mathsf{S}_t,
  \label{eq:D1}
\end{equation}
where $\mathcal{Q}(\mathsf{S}_t)$ denotes the code space stabilized by $\mathsf{S}_t$, and $\mathsf{N}(\mathsf{S}_t)$ is the normalizer of $\mathsf{S}_t$ in the Pauli group. Thus, both the code space and the logical operators evolve dynamically with time. 
Floquet codes form a periodic subclass where the measurement schedule repeats with period $T$:
\begin{equation}
  \mathcal{M}_{t+T}=\mathcal{M}_t.
  \label{eq:D2}
\end{equation}

\begin{table*}[t]
  \centering
  \caption{Comparison of four representative Floquet code families. Geometry indicates the tilings of code construction: Euclidean or (semi‑)hyperbolic tilings.  
The \Encodingrate is \(\lim_{n\to\infty} k/n\).  
Syndrome edges classify whether a single-fault event can flip at most two detectors (\graphedgesyn) or at least three detectors (\hyperedgesyn).}
  \label{tab:floquet_axes_vs_codes}
  \renewcommand{\arraystretch}{1.15}
  \footnotesize
  \begin{tabular}{@{}%
      |p{0.17\linewidth}%  Axis
      |p{0.20\linewidth}%  HCF
      |p{0.20\linewidth}%  HF
      |p{0.17\linewidth}%  FC
      |p{0.13\linewidth}|%  HC
    @{}}
    \toprule
      & \textbf{\mbox{Hyperbolic \kern0.06em color} Floquet code (HCF)} &
        \mbox{Hyperbolic \kern0.06em Floquet} code (HF) &
        Floquet color code &
        Honeycomb \mbox{Floquet \kern0.06em code} \\ \midrule
    \textbf{Geometry} &
      (semi‑)hyperbolic tiling &
      (semi‑)hyperbolic tiling &
      Euclidean tiling &
      Euclidean tiling \\[2pt] \midrule
    \textbf{Typical tiling} &
      \(8.8.8\) (and refinements) &
      \(8.8.8\) (and refinements) &
      \(6.6.6,\;4.8.8\) &
      \(6.6.6,\;4.8.8\) \\[2pt] \midrule
    \textbf{\mbox{Encoding \kern0.06em rate}} \(\displaystyle\lim_{n\to\infty} k/n\) &
      finite (e.g.,\ \(1/8\)) &
      finite (e.g.,\ \(1/8\)) &
      0 &
      0 \\[2pt] \midrule
    \textbf{Measurement schedule} &
      $\XZ{a}{b}$ &
      $\XYZ{a}{b}{c}$ &
      $\XZ{a}{b}$ &
      $\XYZ{a}{b}{c}$ \\[2pt] \midrule
    \textbf{Syndrome edges} &
      \Graphedge &
      \Hyperedge &
      \Graphedge &
      \Hyperedge \\[2pt] \midrule
    \textbf{Effective decoder} &
      \MWPM &
      BP‑based, \BPOSD &
      \MWPM &
      BP-based \\[2pt] \midrule
    \textbf{Threshold} &
      \(\sim\!1.5\,\%\) &
      \(1.5\text{--}2.0\,\%\) \cite{Higgott2024SemiHyper} &
      \(0.3\,\%\) \cite{Kesselring2024},\;\(1.0\,\%\) \cite{Davydova2023} &
      \(1.5\text{--}2.0\,\%\) \cite{Gidney2021Memory} \\ \midrule
    \textbf{Key reference(s)} &
      \textbf{\emph{This work}} &
      \cite{Higgott2024SemiHyper,Fahimniya2024} &
      \cite{Davydova2023,Kesselring2024} &
      \cite{Haah2022Boundaries,Gidney2021Memory} \\ \bottomrule
  \end{tabular}

\end{table*}

%-----------------------------------------------------------------------
\subsection{Euclidean geometry Floquet codes}
\label{sec:landscape-euclid}
%-----------------------------------------------------------------------
Euclidean Floquet codes studied to date share three key structural features:
\textit{(i)} a trivalent, \threecolorable two‑dimensional lattice;  
\textit{(ii)} \weighttwo\ parity‑check measurements; and  
\textit{(iii)} a time‑periodic schedule that realizes a Floquet dynamics.  
These features allow \MWPM\ after embedding the 2D
lattice into a (2{+}1)‑D space‑time graph with only minor changes to
existing algorithms\,\cite{HastingsHaah2021Dynamical}.
In this paper, \(\mathrm{r}\), \(\mathrm{g}\), and \(\mathrm{b}\) denote red, green, and blue edges, respectively (e.g., \(\mathrm{rXX}\) represents an \(X\)-basis parity measurement on red edges).

\paragraph*{Honeycomb Floquet code}
The code measures \(\{X\!\otimes\!X,\,Y\!\otimes\!Y,\,Z\!\otimes\!Z\}\) on the three edge colors of
the hexagonal \(\{6,3\}\) lattice\,\cite{Gidney2021Memory}.  
After \sixstep the
instantaneous stabilizer group (ISG) returns to its initial configuration.  
In this code, the distance scales as \(\dist \propto \sqrt{n}\); on a
torus it grows linearly with code size\,%
\cite{Vuillot2021Planar,Gidney2022Planar,Haah2022Boundaries}.  
Circuit-level simulations under the \EMthree\ noise model give a threshold
\(\simeq 1.5\text{–}2.0\,\%\)\,\cite{Gidney2021Memory}.

\paragraph*{Floquet color code}
The Floquet color code~\cite{Kesselring2024}, also known as the CSS honeycomb 
code~\cite{Davydova2023}, employs a measurement sequence of 
\(\{\mathrm{rXX},\,\mathrm{gZZ},\,\mathrm{bXX},\,\mathrm{rZZ},\,\mathrm{gXX},\,\mathrm{bZZ}\}\).
Importantly, this construction is not obtained by gauge fixing from any parent 
subsystem code~\cite{Davydova2023} and is naturally interpreted as a sequence of
partial anyon condensations in the color code~\cite{Kesselring2024}.
The code distance and encoding rate match those of the honeycomb Floquet code, while the hardware 
requirements are reduced to only two-qubit \(X\) and \(Z\) parity-check modules.
Under circuit-level depolarizing noise, simulations report a threshold of
\(\sim\!0.3\%\)~\cite{Kesselring2024}, with higher values observed under 
biased noise~\cite{setiawan2024}.

%-----------------------------------------------------------------------
\subsection{Hyperbolic geometry Floquet codes}
\label{sec:landscape-hyper}
%-----------------------------------------------------------------------
Negatively curved lattices sacrifice strict planar locality in exchange for improved
resource efficiency~\cite{Breuckmann2017Hyperbolic}. On a closed hyperbolic
surface of genus $\mathfrak{g}$, the first homology group has rank $2\mathfrak{g}$;
therefore, a hyperbolic code can encode $k=2\mathfrak{g}=\Theta(n)$ logical qubits while maintaining $O(1)$ check weight.

\paragraph*{Hyperbolic Floquet (HF) code}
The hyperbolic Floquet code~\cite{Vuillot2021Planar,Higgott2024SemiHyper,Fahimniya2024} 
employs three-colored faces on hyperbolic tilings, with measurements performed on 
the corresponding edge sets according to face colors.
Each round consists of two sub-rounds (creating a three-step period), where 
measurements alternate between the three color classes. After each complete round, 
the ISG returns to its initial form, though 
the geometry of certain logical representatives exhibits a six-step period.
The number of logical qubits scales linearly with physical qubits ($k \propto n$), 
while the distance grows logarithmically ($d \sim \log n$).

\paragraph*{Semi-hyperbolic Floquet code}
Using a fine-graining parameter \(l\), a semi-hyperbolic color code tiling \(T_l\) is constructed from a hyperbolic tiling \(T\) by: (i) taking the dual \(T^*\); (ii) tiling each triangular face of \(T^*\) with a triangular lattice so that each original edge is subdivided into \(l\) edges; and (iii) taking the dual again~\cite{Higgott2024SemiHyper}. The resulting \(T_l\) (e.g., a hexagon–octagon tiling) satisfies \(n_l = l^{2} n\) and \(k_l = k\), and its embedded distance obeys \(d_l \ge C_1\, l\, d\) for a constant \(C_1>0\) depending only on the base hyperbolic tiling. Eliminating \(l\) yields the square-root law \(d_l = \Theta(\sqrt{n_l})\)~\cite{Higgott2024SemiHyper}.

\paragraph*{Hyperbolic color Floquet (HCF) code (this work)}
HCF applies the six-step schedule to hyperbolic tilings such as the regular \(\{8,3\}\) lattice, yielding a finite encoding rate \(k/n=\Theta(1)\) and logarithmic distance \(\dist\sim\log n\). Semi-hyperbolic refinement of the hyperbolic tiling can be combined with the same measurement schedule while preserving the Floquet color code structure and exclusive  \{\XoX, \ZoZ\} \weighttwo\ measurements; in this case \(k\) remains constant and \(\dist\propto\sqrt{n}\).

\paragraph*{Comparative table}\label{par:landscape-table}
Table~\ref{tab:floquet_axes_vs_codes} summarizes the Floquet code families
discussed in Secs.~\ref{sec:landscape-euclid} and \ref{sec:landscape-hyper}.

%-----------------------------------------------------------------------
\subsection{Graph‑edge and Hyperedge Syndromes}
\label{sec:landscape-syndromes}
\begin{figure}[H]
\centering
\includegraphics[width=\linewidth]{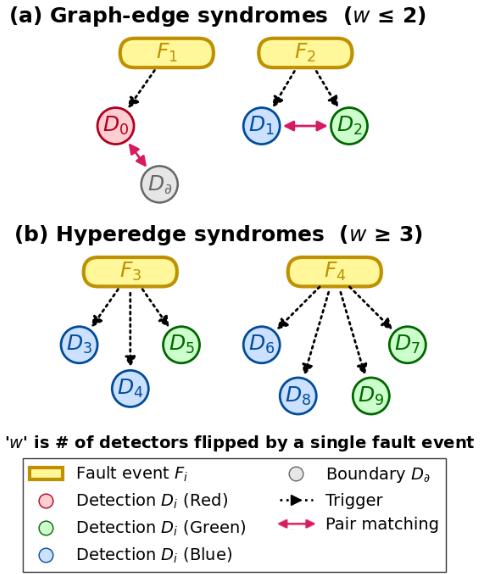}
\caption{Graph‑edge \vs hyperedge syndromes. (a)~\graphedgesyns{} (\(w\!\le2\)).  
A \singlefaultevent triggers at most two detector
nodes, producing a graph that a \MWPM{} decoder can process directly.  
Detector nodes, red, blue, and green circles mark
flips on different stabilizer types; the gray circle
\(D_{\partial}\) is the virtual boundary.  
Black dotted arrows show the triggers from the fault to the detectors.
(b)~\hyperedgesyn{}s (\(w\!\ge 3\)).  
A \singlefaultevent flips three or more detector nodes, producing a hyperedge that connects the detectors.}
\label{fig:graph-hyper-edge}
\end{figure}

%-----------------------------------------------------------------------
Two edge models, the graph-edge and hyperedge syndromes, represent different error structures in quantum error correction. Graph-edge syndromes admit efficient matching-based decoding, while hyperedge syndromes require more sophisticated approaches to handle correlations. We illustrate contrasting syndrome structures in Figure~\ref{fig:graph-hyper-edge}.

\paragraph*{Graph-edge syndromes (\(w\!\le 2\))}
For the case where each \singlefaultevent{} flips at most two detectors (see Figure~\ref{fig:graph-hyper-edge} (a)), the resulting detector graph is the object solved by minimum-weight perfect matching. Edges may be regular (two detectors) or half-edges (one detector to the boundary). In this graph-edge setting, efficient matching decoders achieve empirically near-linear scaling with the number of detectors~\cite{Higgott2025sparseblossom}.

\paragraph*{Hyperedge syndromes ($w\!\ge 3$)}
In contrast, a \hyperedgesyn{} is defined by the presence of a \singlefaultevent{} that flips three or more detector nodes ($w\!\ge 3$), i.e., a fault induces a \emph{hyperedge} connecting more than two detectors, as illustrated in Figure~\ref{fig:graph-hyper-edge}(b). In the setting considered here, parity checks in the three-Pauli basis \{\XoX, \YoY, \ZoZ\} can produce such events via their measurements. Because the resulting structure is not a simple graph, direct application of \MWPM{} to the detector graph is inappropriate. Typical strategies include (i) decomposing each hyperedge into a set of \graphedge{}s, (ii) iteratively reweighting using belief propagation (BP) to capture higher-order correlations, or (iii) employing a decoder that natively handles hypergraph matching.

%======================================================================
\section{Anyon condensation}\label{sec:anyon_overview}
\begin{figure}[H]
  \centering
  \includegraphics[width=0.86\linewidth]{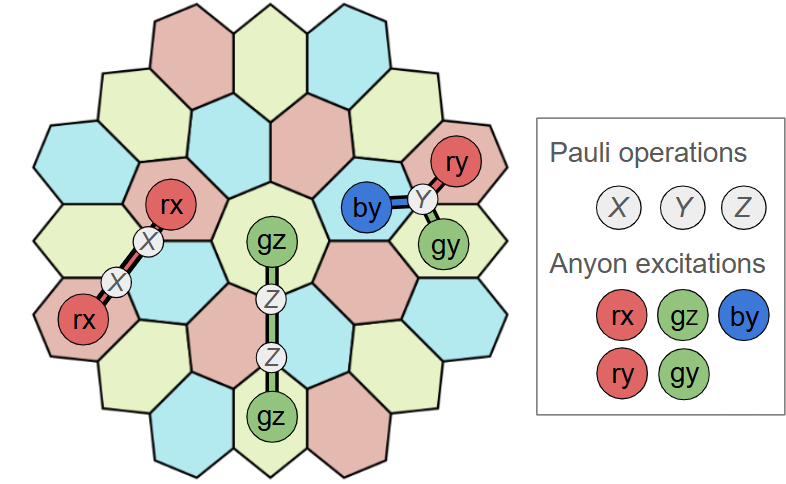}
    \caption{%
      Illustration of anyon generation in the color code at a \semihyperbolic (\(\ell=3\)) lattice.  
      Edge-supported \XoX\ on a red edge (left) produces an
      \(\mathrm{rx}\) anyon pair;  
      \ZoZ\ on a green edge (center) produces a \(\mathrm{gz}\) pair.  
      A single-qubit \(Y\) (right) simultaneously violates the
      three surrounding stabilizers, creating the triplet
      \((\mathrm{ry},\,\mathrm{gy},\,\mathrm{by})\).  
      }
  \label{fig:semihyp_anyon_gen}
\end{figure}

%======================================================================
This section provides a concise overview of anyon excitations \cite{Kitaev2003Anyons} in the color code model (Figure~\ref{fig:semihyp_anyon_gen}) \cite{Bombin2006}, detailing their fusion and braiding rules, as well as their condensation properties~\cite{Kesselring2024}.

%---------------------------------------------------------------------
\subsection{Color code boson table}\label{subsec:fusion_braiding}
%---------------------------------------------------------------------
Following~\cite{Kesselring2018ColorBoundaries}, we introduce \emph{color code boson table}~\eqref{eq:boson_table}. Let $\mathcal{B}_{\mathrm{CC}}=\{c\mu\mid c\in\{\mathrm{r,g,b}\},\ \mu\in\{\mathrm{x,y,z}\}\}$ denote the nine bosons of the color code, which can be arranged in a $3\times3$ table with Pauli rows $(\mathrm{x,y,z})$ and color columns $(\mathrm{r,g,b})$:
\begin{equation}\label{eq:boson_table}
\begin{array}{|c|c|c|}
\hline
\mathrm{rx} & \mathrm{gx} & \mathrm{bx}\\
\hline
\mathrm{ry} & \mathrm{gy} & \mathrm{by}\\
\hline
\mathrm{rz} & \mathrm{gz} & \mathrm{bz}\\
\hline

\end{array}\,.
\end{equation}

\paragraph{Anyon fusion}
All anyons are self-dual, \(a\times a=1\).
Two bosons in \emph{same row} (same Pauli label) or \emph{same column}
(same color) fuse to the third boson in that row/column, 
\eg
\begin{equation}
\mathrm{rx}\times \mathrm{ry}=\mathrm{rz},\qquad
\mathrm{gz}\times \mathrm{bz}=\mathrm{rz}.
\end{equation}

\paragraph{Exchanging and braiding}
The self-exchange (spin) of an anyon $a$ is denoted by $\theta_a\in\{\pm1\}$.
Exchanging two identical Abelian anyons results in a complex phase. 
In qubit stabilizer codes, all elements of $\mathcal{B}_{\mathrm{CC}}$ are bosons, i.e., $\theta_{c\mu}=+1$.

The monodromy (braiding) phase between $a$ and $b$ obeys
\begin{equation}
  M_{a,b}=\frac{\theta_a\,\theta_b}{\theta_{a\times b}}\in\{\pm1\}.
\end{equation}
If $a$ and $b$ share the same Pauli row or the same color column,
then $a\times b$ is the third boson in that row/column and
$M_{a,b}=+1$. If they lie in different rows and columns, then
$a\times b$ is a fermion and $M_{a,b}=-1$.
Examples: \(M_{\mathrm{rx},\mathrm{ry}}=+1\), \(M_{\mathrm{gz},\mathrm{bz}}=+1\), and \(M_{\mathrm{gx},\mathrm{rz}}=-1\).

%---------------------------------------------------------------------
\subsection{Condensation rules}\label{subsec:condense_summary}
%---------------------------------------------------------------------
This section presents condensation rules for color-Pauli anyons. We address three key aspects: the classification of anyons, the construction methodology for instantaneous stabilizer groups, and the characterization of step transitions within the six-step HCF code sequence.

%-----------------------------------------------------------------------
\subsubsection{Classification of anyons}\label{subsec:VDC_rules}
%-----------------------------------------------------------------------
Condensing a boson $a=c\mu\in\mathcal{B}_{\mathrm{CC}}$ corresponds to identifying it with the vacuum, that is, $a\equiv\mathbf{1}$. Under condensation of $a$, the instantaneous stabilizer group (ISG) at that step is denoted by $\mathsf{S}_{a}$.

The protocol for constructing the instantaneous stabilizer group proceeds in two stages. First, checks are added by introducing the shortest hopping operators (open-string moves) that transport the condensed boson $a$. Concretely, this involves inserting the corresponding two-body edge parity checks, such as red-edge \XoX\ checks when $a=\mathrm{rx}$. Second, anti-commuting stabilizers are removed by eliminating any stabilizers that anti-commute with the newly added checks, as these are re-initialized during the subsequent Floquet step.

\begin{figure}[H]
\centering
\includegraphics[width=\linewidth]{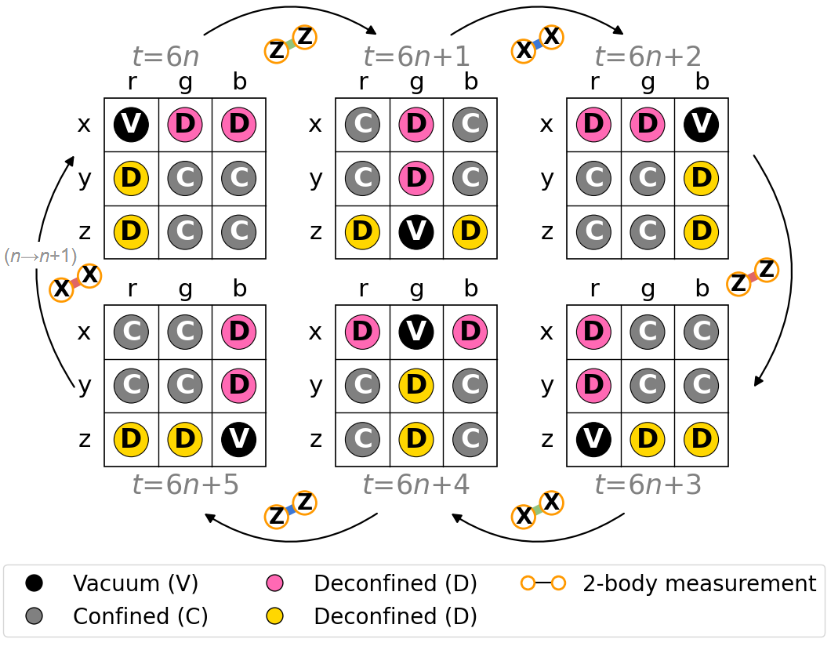}
\caption{%
Evolution of the color-Pauli anyon table.
\(c\mu \bigl(c\!\in\!\{\mathrm{r},\mathrm{g},\mathrm{b}\},\;\mu\!\in\!\{\mathrm{x},\mathrm{y},\mathrm{z}\}\bigr)\)
under the \sixstep \HCF\ code sequence
\(\mathrm{rx}\to\mathrm{gz}\to\mathrm{bx}\to\mathrm{rz}\to\mathrm{gx}\to\mathrm{bz}\).
Each \(3\times 3\) grid is indexed by \textbf{rows}
\((\mathrm{x},\mathrm{y},\mathrm{z}\ \text{Pauli labels})\)
and \textbf{columns} \((\mathrm{r},\mathrm{g},\mathrm{b}\ \text{colors})\).
Cell symbols: \textbf{V}~(black)\,=\,vacuum;
\textbf{C}~(gray)\,=\,confined boson;
\textbf{D}~(pink) and \textbf{D}~(yellow)\,=\,the four deconfined bosons.
The pair of circles with orange outlines joined by a black bar
indicates the two-body \XoX\ or \ZoZ\ parity check
executed in that step; their colors (red/green/blue) indicate the
plaquette involved and the label inside denotes \(X\) or \(Z\).
Black curved arrows show the cyclic progression of the \sixstep.}
\label{fig:boson_color_6step}
\end{figure}

The condensed boson $a$ itself becomes vacuum-identified (\textbf{V}). Any boson sharing either the same row (same Pauli operator) or the same column (same color) as $a$ becomes deconfined (\textbf{D}). In contrast, bosons sharing neither the same row nor the same column with $a$ remain confined (\textbf{C}).
As shown in Figure~\ref{fig:boson_color_6step}, under condensation, the vacuum-identified, deconfined, and confined anyons are marked accordingly on the $3\times3$ table.

%-----------------------------------------------------------------------
\subsubsection{Step transitions}\label{subsec:floquet_transition}
%-----------------------------------------------------------------------
A step $\mathsf{S}_{a}\!\rightarrow\!\mathsf{S}_{b}$ in the Floquet sequence is allowed if and only if $b$ is confined (\textbf{C}) under $\mathsf{S}_{a}$; equivalently, $a$ and $b$ must differ in both color and Pauli labels.
If $b$ is deconfined under $\mathsf{S}_{a}$, then condensing $b$ implements a readout/initialization (logical erasure) rather than a protected transition. Such adjacencies should be avoided in the cycle design.
The six-step \HCF\ code sequence
\begin{equation}
\mathsf{S}_{\mathrm{rx}}\to \mathsf{S}_{\mathrm{gz}}\to \mathsf{S}_{\mathrm{bx}}\to
\mathsf{S}_{\mathrm{rz}}\to \mathsf{S}_{\mathrm{gx}}\to \mathsf{S}_{\mathrm{bz}}
\end{equation}
satisfies compatibility at each adjacent step, as illustrated in Figure~\ref{fig:boson_color_6step}.

%======================================================================
\section{Code construction}
\label{sec:code_construction}
%======================================================================
In this section, we describe the construction of detectors and logical operators for \HCF\ code. Throughout this paper, one round is defined as the six-step Floquet schedule.

%-----------------------------------------------------------------------
\subsection{Detector structure}
\label{sec:detector_3d}
%-----------------------------------------------------------------------

At the Floquet periodic schedule, an instantaneous stabilizer group $\mathsf{S}_a$ is obtained by condensing a color--Pauli boson $a$. 
Moreover, $\mathsf{S}_a$ is generated by \emph{all} plaquette stabilizers together with the weight-two edge checks associated with the condensed color. 
Since plaquette stabilizers commute with every edge check in the schedule, their eigenvalues remain well defined throughout the sequence. 
Let $S^{(c,\mu)}_{f}$ denote the face-supported plaquette stabilizers, with $c\in\{\mathrm{r},\mathrm{g},\mathrm{b}\}$, $\mu\in\{\mathrm{x},\mathrm{y},\mathrm{z}\}$, and $f$ a face index.
These generators mutually commute and also commute with all weight-two edge checks used in the measurement schedule.
During a Floquet step that measures weight-two edge checks of a given color, the system remains in an eigenstate of all plaquette stabilizers on the other two colors; in particular, the value of $S^{(c,\mu)}_{f}$ can be \emph{inferred} from the outcomes of an internally commuting measurement set for that step.
These inferred values at two instants are used to define detector bits below.

%---------------------------------------------------------------------
\subsubsection{Detector construction in \HF\ code}
\label{subsubsec:hf-recap}
%-----------------------------------------------------------------------

\paragraph{Consecutive \Nstep{two} checks}
\label{subsubsec:time-overlap}

In \HF\ code, the \twobody measurement schedule is arranged  so that detector value is distributed across two consecutive steps. For a fixed color \(c\) and Pauli type \(\mu\), taking the product of those two steps yields a measurement of the plaquette stabilizer $S^{(c,\mu)}_{f}$. Within one period of the schedule, we designate a
\emph{leading} \Nstep{two} checks of consecutive steps and a \emph{trailing} \Nstep{two} checks of consecutive steps; For a fixed color \(c\) and Pauli type \(\mu\), each \Nstep{two} checks produce a separate inference of the detector value for \((c,\mu)\).

\begin{figure}[H]
  \centering
  \includegraphics[width=\linewidth]{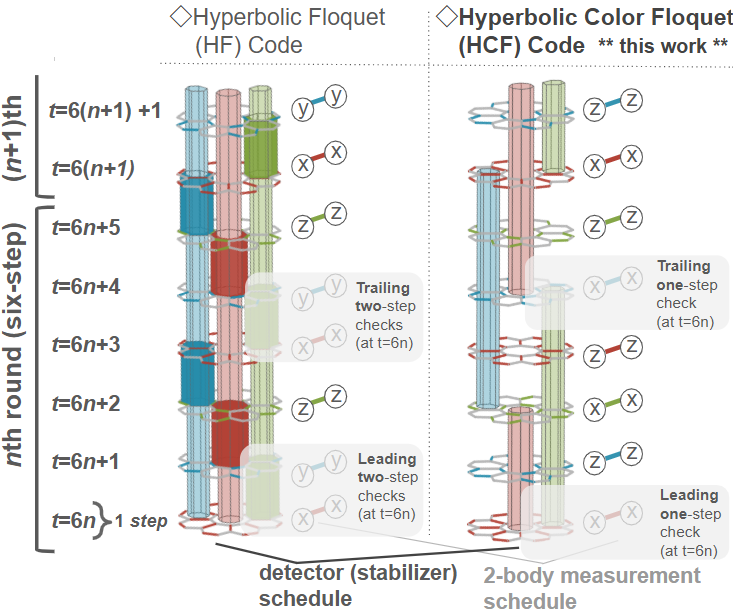}
  \caption[Detector pillars in 3D spacetime]{%
    Detector pillars in \protect\threeD\ spacetime.\par
    A ``detector pillar'' is a visualization formed by connecting, along the time axis, the two plaquettes associated with a stabilizer (a detector) measured at two corresponding times.\par
    \emph{Left (\protect\HF\ code).} The schedule of detectors consists of a \emph{leading \protect\Nstep{two} checks} and a \emph{trailing \protect\Nstep{two} checks}.
    Each plaquette value is inferred as the product of two consecutive steps within the corresponding pair, thus one detector requires four measurement steps in total: two for the leading inference and two for the trailing inference.
    In the pillar representation, the two thick segments correspond to these two products (leading and trailing). The darker, thicker appearance emphasizes that, the adjacent stabilizers share the measurements.\par
    \emph{Right (\protect\HCF\ code).} The measurement schedule partitions edges by color, allowing a plaquette value to be inferred within a \emph{\protect\Nstep{one} check}.
    Accordingly, a detector consists of two \protect\Nstep{one} check of the plaquette stabilizer $S^{(c,\mu)}_{f}$ taken at different times (two measurement steps in total).
    For simplicity, we visualize only the detector pillars associated with Pauli-$X$ checks; the Pauli-$Z$ based detectors are not shown.%
  }
  \label{fig:detector-pillars}
\end{figure}

The left panel of Figure~\ref{fig:detector-pillars} shows the
spacetime support: the two thick (dark) segments correspond to the leading
and trailing \Nstep{two} checks, respectively. 
Let \(\Pi_c^\mu[t:t{+}1]\in\{0,1\}\) denote the \Nstep{two} estimate of the
\((c,\mu)\) detector value formed by taking the product—using additive notation in
\(\mathbb{F}_2\) and multiplicative notation in \(\{\pm1\}\), with \(\oplus\) denoting addition
modulo \(2\)—of the commuting checks collected at steps \(t\) and \(t{+}1\).
Choose a \emph{leading} pair \([t_{\mathrm{L}},t_{\mathrm{L}}{+}1]\) and a \emph{trailing} pair \([t_{\mathrm{T}},t_{\mathrm{T}}{+}1]\) with
\(t_{\mathrm{L}}{+}1 < t_{\mathrm{T}}\). The detector compares these two
pairwise inferences:
\begin{equation}
    d_{\HF}^{(c,\mu)} \;=\;
    \Pi_c^\mu\!\big[t_{\mathrm{L}}:t_{\mathrm{L}}{+}1\big]
    \;\oplus\;
    \Pi_c^\mu\!\big[t_{\mathrm{T}}:t_{\mathrm{T}}{+}1\big].
    \label{eq:hf-detector}
\end{equation}

\paragraph{Time-overlapping detector}
\label{subsubsec:time-overlap}
In \HF\ code, detectors are arranged so that temporally 
adjacent detectors share a subset of measurement checks.
We refer to this structure as the \emph{time-overlapping} detector construction.
For the periodic schedule shown in Figure~\ref{fig:detector-pillars} (with color \(c\) and Pauli type \(\mu\) fixed), each detector is formed from a pair consisting of a \emph{leading} two-step inference and a \emph{trailing} two-step inference, namely \([t_\mathrm{L}, t_\mathrm{L}{+}1]\) and \([t_\mathrm{T}, t_\mathrm{T}{+}1]\) with \(t_\mathrm{L}{+}1 < t_\mathrm{T}\).
More concretely, for an integer \(n\in\mathbb{Z}\), define
\begin{equation}
\begin{split}
\mathcal{C}^{\mathrm{lead}}_{n} &= \{6n,\,6n{+}1\},\\
\mathcal{C}^{\mathrm{trail}}_{n} &= \{6n{+}3,\,6n{+}4\}.
\end{split}
\end{equation}
Then the (green) \emph{earlier-in-time} detector, regarded as centered at \(t=6n{+}4\), uses the four checks
\begin{equation}
\begin{split}
\mathcal{D}_{n}
&= \mathcal{C}^{\mathrm{lead}}_{n}\;\cup\;\mathcal{C}^{\mathrm{trail}}_{n}\\
&= \{6n,\,6n{+}1,\,6n{+}3,\,6n{+}4\}.
\end{split}
\end{equation}
The subsequent (green) \emph{later-in-time} detector, regarded as centered at \(t=6(n{+}1){+}1\), uses
\begin{equation}
\begin{split}
\mathcal{D}_{n{+}1}
&= \mathcal{C}^{\mathrm{trail}}_{n}\;\cup\;\mathcal{C}^{\mathrm{lead}}_{n{+}1}\\
&= \{6n{+}3,\,6n{+}4,\,6(n{+}1),\,6(n{+}1){+}1\}.
\end{split}
\end{equation}

Therefore, the two adjacent detectors share
\begin{equation}
\begin{split}
\mathcal{D}_{n}\cap\mathcal{D}_{n{+}1}
&= \mathcal{C}^{\mathrm{trail}}_{n}\\
&= \{6n{+}3,\,6n{+}4\},
\end{split}
\end{equation}
which is precisely the \emph{time-overlap}.

%---------------------------------------------------------------------
\subsubsection{Detector construction in \HCF\ code}
\label{subsubsec:hcf-recap}
%-----------------------------------------------------------------------

\paragraph{One-step checks}
%\label{subsubsec:hcf-time-gap}
In \HCF\ code, trailing and leading check measurements are extracted through \emph{one-step checks}, where the detectors are then constructed by comparing these one-step checks across different time steps.
Formally, for a plaquette of color $c$ and Pauli type $\mu \in \{X,Z\}$, let $P_c^\mu(t) \in \{0,1\}$ denote the one-step estimate obtained from measurements at time $t$. A detector compares the estimates of the initial and final times $t_{\mathrm{i}}$ and $t_{\mathrm{f}} > t_{\mathrm{i}}$:
\begin{equation}
    d_{\mathrm{HCF}}^{(c,\mu)} = P_c^\mu(t_{\mathrm{i}}) \oplus P_c^\alpha(t_{\mathrm{f}}),
    \label{eq:hcf-detector}
\end{equation}
where $\oplus$ denotes the addition in $\mathbb{F}_2$.

\paragraph{Time-gapped detector}
\label{subsubsec:hcf-time-gap}
In \HCF\ code, temporally adjacent detectors of the same color do \emph{not} reuse the same check values.
We refer to this structure as the \emph{time-gapped} detector construction.
As illustrated in the right panel of Figure~\ref{fig:detector-pillars} for the six-step schedule, the (green) \emph{earlier-in-time} detector, regarded centered at \(t=6n{+}4\), compares the \Nstep{one} inferences at times \(t=6n\) and \(t=6n{+}4\), and thus uses the two check groups.
\begin{equation}
\mathcal{D}_{n}=\{6n,\,6n{+}4\}.
\end{equation}
The subsequent (green) \emph{later-in-time} detector, regarded as centered at \(t=6(n{+}1){+}4\), compares the \Nstep{one} inferences at \(t=6(n{+}1)\) and \(t=6(n{+}1){+}4\), using
\begin{equation}
\mathcal{D}_{n{+}1}=\{6(n{+}1),\,6(n{+}1){+}4\}.
\end{equation}
Consequently, adjacent detectors do not share checks,
\begin{equation}
\mathcal{D}_{n}\cap\mathcal{D}_{n{+}1}=\varnothing,
\end{equation}
and there is a one-step temporal gap at \(t=6n{+}5\) between their supports; this is precisely the \emph{time-gapped}.

%-----------------------------------------------------------------------
\subsection{Logical operator}
\label{sec:static_loops}
%-----------------------------------------------------------------------

\begin{figure}[H]
  \centering
  \includegraphics[width=\linewidth]{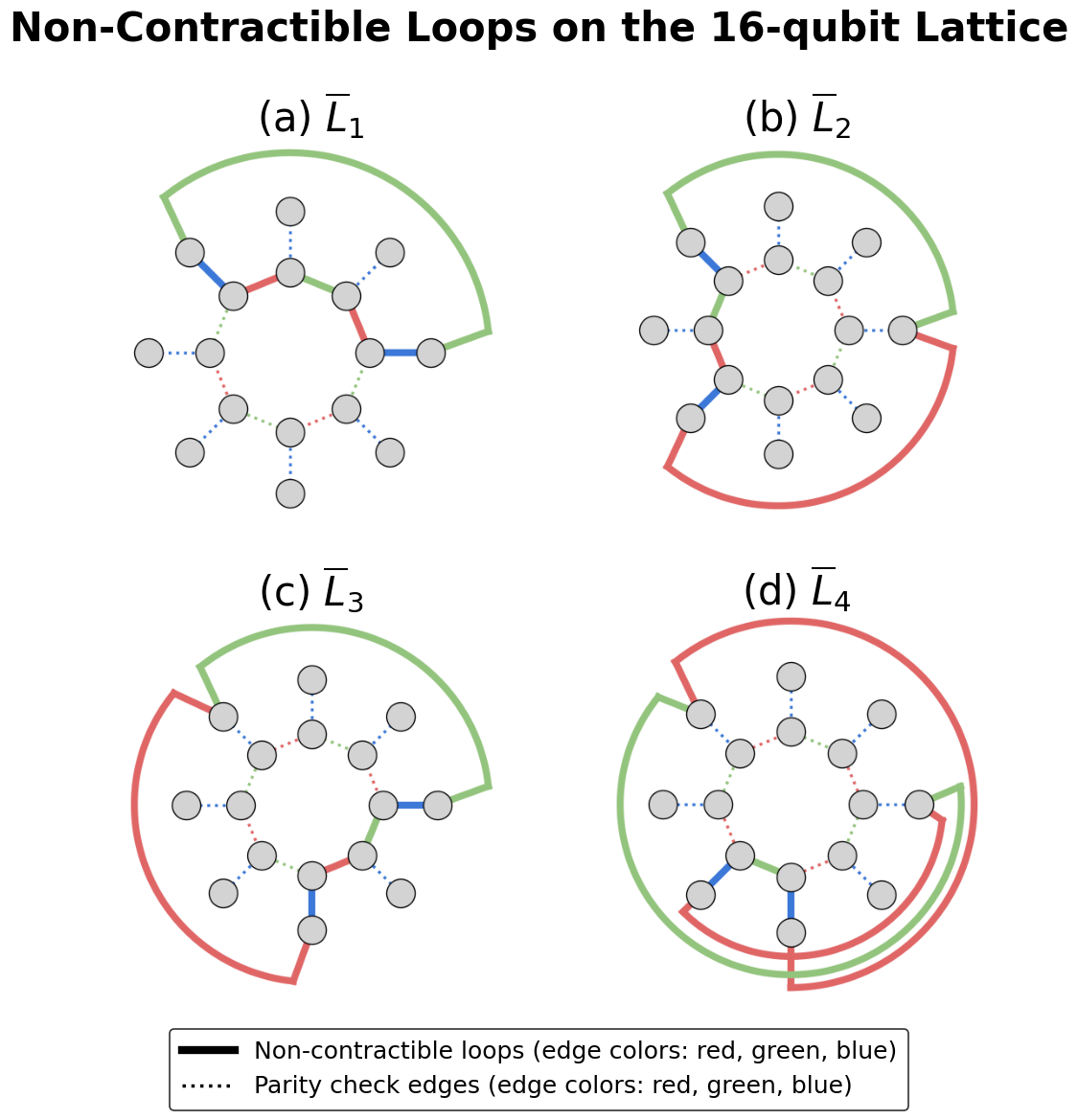}
  \caption{Non-contractible loops on the 16-qubit lattice. Panels (a)-(d) 
  display non-contractible closed paths $\gamma_1$-$\gamma_4$. Thick solid 
  arcs show the supports of loop operators $\bar{L}_i \equiv L(\gamma_i)$, 
  dotted segments indicate parity-check edges, and colors (red/green/blue) 
  correspond to the three check types.}
  \label{fig:H16_loops_static}
\end{figure}

\paragraph*{Non-contractible loop}
The 16-qubit lattice forms a closed genus-2 surface with $|V|=16$ vertices 
and $|E|=24$ edges, containing $2\mathfrak{g}=4$ linearly independent, homologically 
non-contractible loops $\gamma_i$ (Figure~\ref{fig:H16_loops_static} shows 
four representative examples). 

For a closed path $\gamma$ in the lattice, 
the loop operator is defined as
\begin{equation}
L(\gamma) \coloneqq \prod_{e\in\gamma} O_e,
\end{equation}
where $O_e$ denotes the local operator on edge $e$. If $L(\gamma)$ can be 
expressed as a product of plaquette operators, then $L(\gamma)$ is contractible.

Each loop is represented by the cyclic sequence
\begin{equation}
\gamma_i = (v^{(i)}_0, \ldots, v^{(i)}_{\ell_i-1}),
\end{equation}
with edges $e^{(i)}_k = (v^{(i)}_k, v^{(i)}_{k+1})$ and 
$v^{(i)}_{\ell_i} \equiv v^{(i)}_0$. 
The $\mathbb{Z}_2$ intersection form
\begin{equation}
  \langle[\gamma_i],[\gamma_j]\rangle = \#(\gamma_i \cap \gamma_j) \bmod 2
\end{equation}
is non-singular over $\mathbb{F}_2$. Therefore, the homology classes 
$[\gamma_i] \in H_1(\Sigma; \mathbb{Z}_2)$ serve as a homology basis 
that generates—up to stabilizers—the space of logical operators.

\noindent\textbf{Genus-2 example.}
For a closed surface of genus $\mathfrak{g}=2$, the first homology group with 
$\mathbb{Z}_2$ coefficients is $H_1(\Sigma_2;\mathbb{Z}_2) \cong 
\mathbb{Z}_2^{2g} = \mathbb{Z}_2^4$. For a standard symplectic basis 
$\{[\alpha_1], [\beta_1], [\alpha_2], [\beta_2]\}$ with 
$\langle[\alpha_i], [\beta_j]\rangle = \delta_{ij}$ and all other 
pairings zero, the $\mathbb{Z}_2$ intersection matrix is
\begin{equation}
  J =
  \begin{pmatrix}
    0 & 1 & 0 & 0\\
    1 & 0 & 0 & 0\\
    0 & 0 & 0 & 1\\
    0 & 0 & 1 & 0
  \end{pmatrix},
  \qquad
  \det(J) \equiv 1 \pmod{2}.
\end{equation}
Since the pairing is non-singular, the four non-contractible classes 
$[\alpha_1], [\beta_1], [\alpha_2], [\beta_2]$ form a homology basis. 
Loop operators supported on these curves generate—modulo stabilizers—the 
logical operator space.

\paragraph*{Construction of logical loop}
\label{subsec:logical_loops_construction}

For each $\beta \in \{X,Z\}$, a logical loop representative 
$\bar{L}^{\beta}_t(\gamma) \in \mathcal{P}_n$ is defined at time $t$. 
The support of $\bar{L}^{\beta}_t(\gamma)$ is restricted to $\gamma$; 
only its Pauli type and frame can change with $t$. The objective is to 
update $\bar{L}^{\beta}_t(\gamma)$ after each step to ensure commutation 
with the subsequent measured checks $\mathcal{M}_{t+1}$.

\begin{figure}[H]
  \centering
  \includegraphics[width=\linewidth]{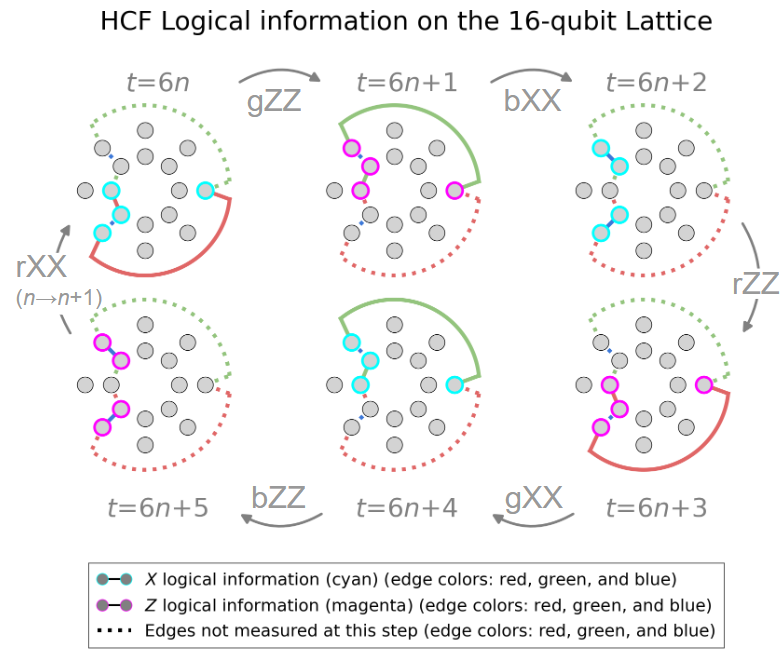}
  \caption{Logical information of $\bar{L}_2$ through Floquet period on the 16-qubit lattice. The measurement 
  sequence of HCF code follows \(\{\mathrm{rXX},\,\mathrm{gZZ},\,\mathrm{bXX},\,\mathrm{rZZ},\,\mathrm{gXX},\,\mathrm{bZZ}\}\). Solid 
  edges indicate current measurements; dotted edges show unmeasured 
  interactions. Highlighted nodes indicate the logical 
  operator type: cyan for X-type, magenta for Z-type. Grey arrows 
  show return to initial configuration after six steps.}
  \label{fig:HCF_logical_operator_cycle}
\end{figure}

Let $c_t \in \{\mathrm{r}, \mathrm{g}, \mathrm{b}\}$ denote the color 
measured in time $t$, and let $E_{c_t}$ be the set of edges of that color. 
Let $m_e(t) \in \{\pm 1\}$ denote the outcome of the weight-two check on 
edge $e$ at time $t$. The loop update operator is defined as the product 
of checks on $\gamma$ that were just measured:
\begin{equation}
    \begin{aligned}
        U_t(\gamma)
        &\coloneqq
        \prod_{e \in \gamma \cap E_{c_t}} \mathsf{C}^{(t)}_e,\\[4pt]
        \mathsf{C}^{(t)}_e
        &\in
        \begin{cases}
        \{\mathrm{XX}_e, \mathrm{YY}_e, \mathrm{ZZ}_e\} & \text{(HF code)}\\[2pt]
        \{\mathrm{XX}_e, \mathrm{ZZ}_e\}      & \text{(HCF code)}
        \end{cases}.
    \end{aligned}
\end{equation}

The logical loop representative is then updated (see also Figure~\ref{fig:HCF_logical_operator_cycle}) by
\begin{equation}
  \begin{aligned}
  \bar{L}^{\beta}_{t+1}(\gamma)
  &= U_t(\gamma) \bar{L}^{\beta}_t(\gamma),\\
  \beta &\in \{X,Z\}.
  \end{aligned}
  \label{eq:logical-loop-update}
\end{equation}
With this update, the representative $\bar{L}^{\beta}_{t+1}(\gamma)$ commutes with the checks scheduled in the following step $\left(\mathcal{M}_{t+1}\right)$.

The corresponding Pauli frame accumulates the measured signs along $\gamma$:
\begin{equation}
\label{eq:logical-loop-frame}
\begin{aligned}
  \sigma^{\beta}_{t+1}(\gamma)
  &= \sigma^{\beta}_{t}(\gamma) \cdot
     \prod_{e \in \gamma \cap E_{c_t}} m_e(t),\\
  \sigma^{\beta}_t(\gamma)
  &\in \{\pm 1\}.
\end{aligned}
\end{equation}

%======================================================================
\section{Noise model and decoder}
\label{sec:noise_decoder}
%======================================================================

In this section, we describe the noise models used to evaluate the performance and the decoders used to correct errors. We focus on circuit-level noise models, including both independent and correlated error scenarios.

\begin{table}[H]
\centering
\footnotesize
\setlength{\tabcolsep}{4pt}
\renewcommand{\arraystretch}{1.1}
\setlength{\arrayrulewidth}{0.6pt} % 枠線の太さ
\caption{EM3 noisy gates  (following~\cite{Gidney2022Planar})}
\label{tab:em3_gates}
\begin{tabular}{|p{0.16\columnwidth}|p{0.44\columnwidth}|p{0.30\columnwidth}|}
\hline
\textbf{Noise} & \textbf{\EMind} & \textbf{\EMcor} \\
\hline
Clifford$_1$ &
1-qubit Clifford $+$ depolarizing$_{1\mathrm{q}}$ (prob.\ $p$) &
(not used) \\
\hline
Clifford$_2$ &
2-qubit Clifford $+$ depolarizing$_{2\mathrm{q}}$ (prob.\ $p$) &
(not used) \\
\hline
Prepare$_P$ &
Prepare $+1$ eigenstate of $P\!\in\!\{X,Z\}$; flip (prob.\ $p$) &
(prob.\ $p/2$) \\
\hline
Meas.$_{PP}$ (2q parity) &
\emph{Independent} measurement flip (prob.\ $p$) &
\emph{Correlated} with prob.\ $p$, $\{I,X,Y,Z\}^{\otimes 2}\!\times\!\{\mathrm{flip},\mathrm{no\ flip}\}$ \\
\hline
Idle &
depolarizing$_{1\mathrm{q}}$ (prob.\ $p$) &
depolarizing$_{1\mathrm{q}}$ (prob.\ $p$) \\
\hline
\end{tabular}
\end{table}

%---------------------------------------------------------------------
\subsection{Noise model}
\label{subsec:noise_overview}
%---------------------------------------------------------------------
Following~\cite{Gidney2022Planar}, two noise models are considered that differ only 
in how the error budget is allocated across different noisy gates 
(summarized in Table~\ref{tab:em3_gates}):
\begin{itemize}[leftmargin=1.6em]
  \item \textbf{Independent entangling-measurement noise (\EMind):} Every gate fails independently with probability $p$,
  modeled as follows: a single-qubit (two-qubit) Clifford is followed by a single-qubit 
  (two-qubit) depolarizing channel with probability $p$; preparing the $+1$
  eigenstate of $P \in \{X,Z\}$ results in a flip to the $-1$ eigenstate with 
  probability $p$; a two-qubit $PP$ parity measurement is preceded by an uncorrelated
  two-qubit depolarizing channel with probability $p$; and idle operations undergo
  single-qubit depolarizing with probability $p$.

\item \textbf{Correlated entangling-measurement noise (\EMcor):} This error model focuses on parity measurement operations. With probability $p$, correlated error events apply a pair of Pauli operators from $\{I,X,Y,Z\}^{\otimes 2}$ and affect the measurement outcome $\{\mathrm{flip},\mathrm{no\ flip}\}$. Single- and two-qubit Clifford gates remain noiseless, preparation errors occur with reduced probability $p/2$, and idle operations undergo single-qubit depolarizing noise with probability $p$.

\end{itemize}

%-----------------------------------------------------------------------
\subsection{Decoder}
\label{subsec:decoder_overview}
%-----------------------------------------------------------------------

We briefly summarize the two decoders used in this work.

\paragraph{MWPM (Minimum-Weight Perfect Matching)} 
For surface codes, $X$- and $Z$-type syndromes are processed in
separate instances of a pairing problem on a detector-level
matching graph \cite{Dennis2002Topological,Higgott2025sparseblossom,Fowler2012Surface}.
Nodes are detectors (and boundaries), and edges
encode \graphedge error mechanisms that flip one or two detectors.
Edge weights are derived from a prior noise model (e.g.,\ $w(e)=-\log
p(e)$), and a minimum-total-weight perfect matching is computed, then realized along weighted-shortest paths to synthesize the recovery. 

\paragraph{BP+OSD (Belief Propagation + Ordered Statistics Decoding)} 
Belief propagation (BP) is run on the circuit/code Tanner graph to estimate posterior log-likelihood ratios for error mechanisms (variable nodes). Because cycles in the Tanner graph and degeneracy can stall BP, an ordered statistics decoder (OSD) augments it by selecting a most-reliable basis and enumerating test patterns up to a specified order among the least-reliable bits, reconstructing valid candidates and selecting the maximum-likelihood solution \cite{Panteleev2021,roffe2020BPOSD}.
BP operates in linear time per iteration, whereas the OSD stage dominates the computational complexity and scales with the chosen order, yielding a tunable accuracy–runtime trade-off. This approach is broadly applicable to sparse-graph quantum codes beyond planar surface codes.

%======================================================================
\section{Numerical calculation}
\label{sec:calculation}
%======================================================================
The following four subsections present a numerical study of \HF\ and \HCF\ code families.
%----------------------------------------------------------------------
\subsection{Impact of noise models and code sizes on performance}
\label{subsec:noise_size}
%-----------------------------------------------------------------------
Performance evaluation investigates how different noise characteristics 
and code sizes influence the logical error rates of HF and HCF codes. 

For each physical error rate $P_{\mathrm{phys}}$ (equivalent to the 
probability $p$ in Section~\ref{subsec:noise_overview}), we execute 
48 time steps with either EM-independent or EM-correlated noise. 
Syndromes are decoded using minimum-weight perfect matching 
(MWPM). We consider three code sizes $n \in \{16, 64, 144\}$ with 
corresponding distances $d \in \{2, 3, 4\}$, respectively.

\begin{figure}[H]
  \centering
  \includegraphics[width=\linewidth]{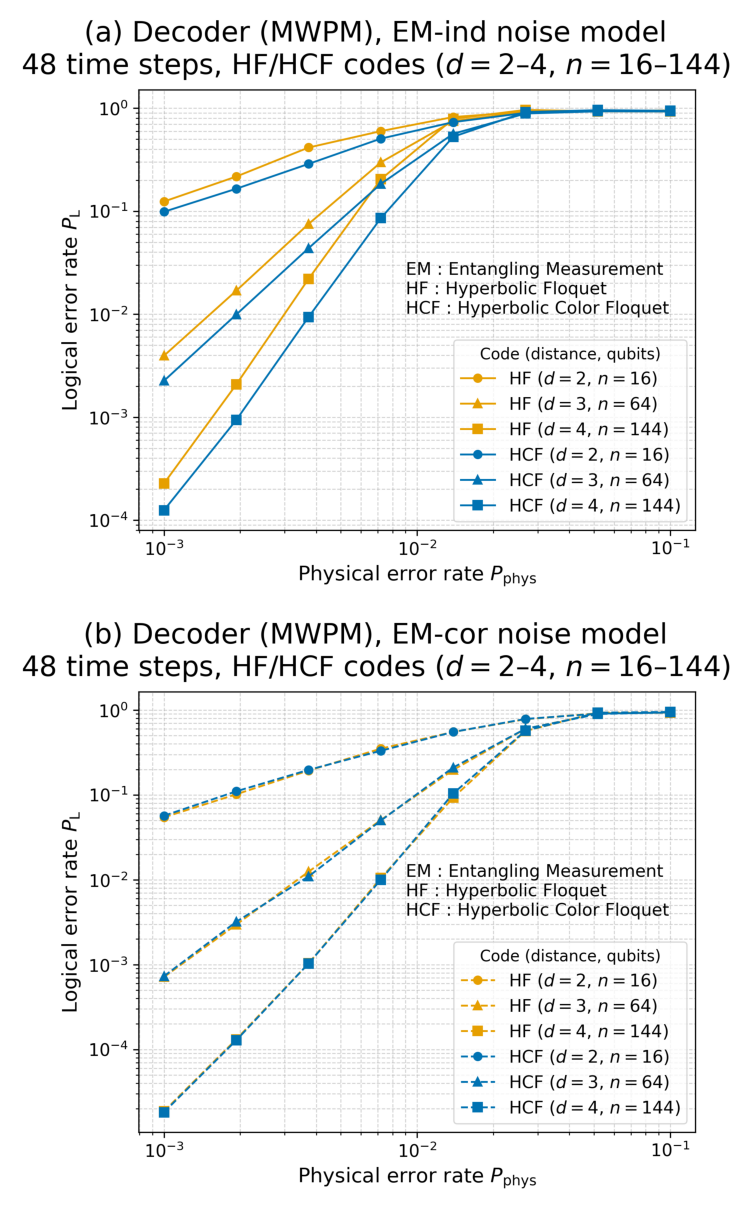}
  \caption{Comparison of logical error rates between HF and HCF codes. (a) EM-independent noise. (b) EM-correlated 
  noise. Curves show logical error rate \vs physical error rate for 
  HF code (orange) and HCF code (blue). Markers indicate code sizes $n=16, 64, 144$ 
  (distances $d=2, 3, 4$). In panel (a) at $d=4$ and $P_{\mathrm{phys}}=3.7 \times 10^{-3}$, 
  HF yields $P_{\mathrm{L}}=2.2 \times 10^{-2}$ while HCF code achieves $9.4 \times 10^{-3}$.}
  \label{fig:HF_vs_HCF_noise}
\end{figure}

Figure~\ref{fig:HF_vs_HCF_noise} compares the logical error rates for the HF code (orange) and HCF code (blue). Across almost all sizes and both noise models, HCF code maintains lower logical error rates compared with the HF code; at $d=4$ and $P_{\mathrm{phys}}=3.7 \times 10^{-3}$, the reduction reaches a factor of approximately 2.3 at Figure~\ref{fig:HF_vs_HCF_noise} (a).

%----------------------------------------------------------------------
\subsection{Logical performance across decoders}
\label{subsec:dec}
%----------------------------------------------------------------------

\begin{figure}[H]
  \centering
  \includegraphics[width=\linewidth]{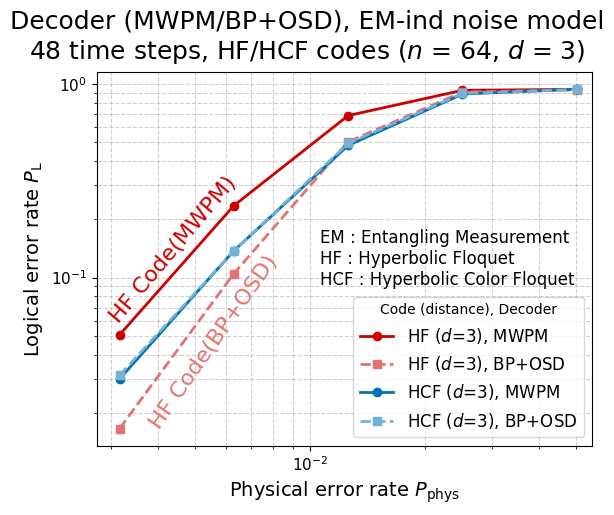}
  \caption{Impact of decoder choice on HF and HCF codes at $n=64$ under 
  EM-independent noise. Solid lines represent MWPM; dashed lines represent 
  BP+OSD (order 1). At $P_{\mathrm{phys}}=3.2 \times 10^{-3}$ (markers), HCF code achieves 
  $P_{\mathrm{L}}=3.0 \times 10^{-2}$ (MWPM) and $3.1 \times 10^{-2}$ (BP+OSD), 
  corresponding to a 4\% variation. HF code yields $P_{\mathrm{L}}=5.1 \times 10^{-2}$ 
  (MWPM) and $1.7 \times 10^{-2}$ (BP+OSD), demonstrating a threefold 
  reduction in error rate. Graph-edge syndromes in HCF code are well-matched to MWPM, 
  while hyperedge syndromes in HF code benefit from advanced decoding.}
  \label{fig:decoder_impact_64}
\end{figure}

With the code size fixed at $n=64$, this section examines how decoder choice affects logical performance. 
Numerical simulations under EM-independent noise were conducted to compare MWPM and BP+OSD decoders, 
as illustrated in Figure~\ref{fig:decoder_impact_64}.

At $P_{\mathrm{phys}}=3.2 \times 10^{-3}$, the decoder improvement ratio 
$\Delta_{\mathrm{dec}} = P_{\mathrm{L}}^{\mathrm{MWPM}}/P_{\mathrm{L}}^{\mathrm{BP+OSD}}$ equals 
3.0 for HF code but $9.6 \times 10^{-1}$ for HCF code, demonstrating that HF code benefits 
substantially from BP+OSD while HCF code shows negligible difference. 
This disparity arises because BP+OSD effectively handles the high-degree 
hyperedge syndromes that challenge MWPM in HF code, explaining the performance gain.

This trend in HF and HCF codes aligns with Euclidean results 
for honeycomb Floquet and Floquet color codes: under correlated noise, BP-assisted matching outperforms 
standard MWPM~\cite{DerksEtAl2025NoisyReadouts}.

In addition to logical performance, we report numerical evaluations of the decoding time in Appendix~\ref{app:decoding_time}

%---------------------------------------------------------------------
\subsection{Detector weight distribution}
\label{subsec:detector_weight_results}
%-----------------------------------------------------------------------

This subsection compares the detector weight distribution $w$ (defined as the number of detector nodes triggered by a single-fault event) between HF and HCF codes. The analysis uses circuit-level \EMind\ noise model. 
HF code employs a time-overlapping construction with three-Pauli basis 
$\{\XoX, \YoY, \ZoZ\}$, while HCF code uses time-gapped construction with 
two-Pauli basis $\{\XoX,\ZoZ\}$, comparing the same plaquette at times $t$ 
and $t+4$ (4-step interval). Details of the schedule and detector 
constructions are provided in the appendix~\ref{app:detector_weight}.

\begin{figure}[H]
\centering
\includegraphics[width=\linewidth]{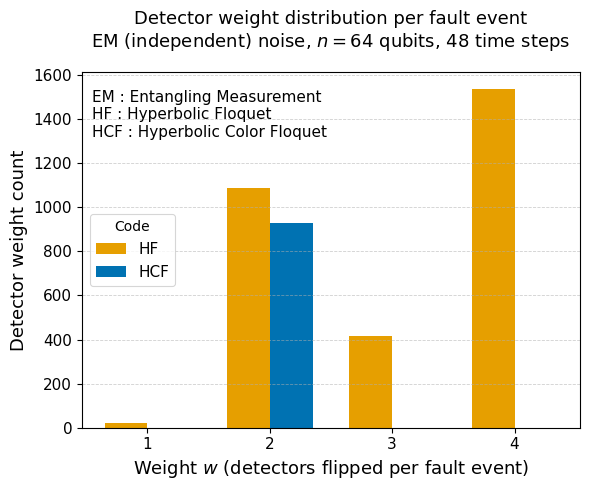}
  \caption{Detector-weight distribution (\HF\ vs.\ \HCF codes ; \EMind\ noise model, code size \(n=64\), 48 time steps).
  Here \(w\) is the number of detectors flipped per a \singlefaultevent.
  \HF\ code shows both even (\(w{=}2,4\)) and odd (\(w{=}1,3\)) weights, whereas \HCF\ code is concentrated at \(w{=}2\).}
\label{fig:detector_weight_hist}
\end{figure}

Figure~\ref{fig:detector_weight_hist} and Table~\ref{tab:fault-degree-stats} reveal two distinct patterns under \EMind\ noise model with \(n=64\) over 48 time steps (here \(w\) denotes the number of detectors flipped per a \singlefaultevent).
(i) In \HF\ code, the distribution is spread over even and odd weights: \(w{=}4\) accounts for 50.2\%, \(w{=}2\) for 35.6\%, and odd weights \(w{=}1,3\) collectively for 14.3\%.
(ii) In \HCF\ code, the weights concentrate entirely at \(w{=}2\), with no occurrences at $w = 1, 3, \text{ or } 4$.

For the mechanisms by which single-fault events in the HF and HCF codes produce their respective detector weight patterns , see Appendix~\ref{app:detector_weight}.

% -- Table: detector-weight statistics -------------------------------
\begin{table}[H]
  \centering
  \caption{Detector weight statistics over 48-time steps
           under \EMind noise model ($n=64$).}
  \label{tab:fault-degree-stats}
  \renewcommand{\arraystretch}{1.1}
  \footnotesize
  \begin{tabular}{@{}c r r r r@{}}
    \toprule
        & \multicolumn{2}{c}{\HF} & \multicolumn{2}{c}{\HCF} \\[-0.2ex]
    \cmidrule(lr){2-3}\cmidrule(l){4-5}
    $w$ & count & \% & count & \% \\
    \midrule
     1 &   20 &  0.7 &   0 &   0.0 \\
     2 & 1088 & 35.6 & 928 & 100.0 \\
     3 &  416 & 13.6 &   0 &   0.0 \\
     4 & 1536 & 50.2 &   0 &   0.0 \\
    \midrule
    $\Sigma$ & 3060 &      & 928 &        \\
    \bottomrule
  \end{tabular}
\end{table}

%---------------------------------------------------------------------
\subsection{Circuit-level threshold}\label{subsec:threshold}
%-----------------------------------------------------------------------

HCF code circuits are simulated for a single six-step Floquet period 
under the EM-independent noise model. The corresponding code distances are $d = 3, 4, 7, 11$.

\begin{figure}[H]
\centering
\includegraphics[width=\linewidth]{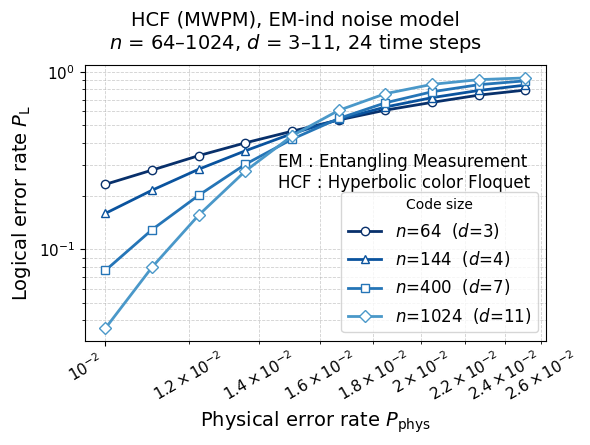}
\caption{Circuit-level threshold of HCF code under EM-ind noise. 
Log-log plot of logical error rate \vs physical error rate $P_{\mathrm{phys}}$ for 
HCF code with $n = \{64, 144, 400, 1024\}$ corresponding to distances 
$d = \{3, 4, 7, 11\}$. All curves intersect at $P_{\mathrm{phys}} \approx 1.4$–$1.6\%$.}
\label{fig:thres_hcf}
\end{figure}

Figure~\ref{fig:thres_hcf} shows all curves crossing within the narrow interval $P_{\mathrm{phys}} \simeq 1.4$–$1.6\%$. The circuit-level threshold is therefore approximately 1.5\%.
This HCF code's threshold falls within the $1.5$–$2.0\%$ range reported for HF code~\cite{Higgott2024SemiHyper}.

%======================================================================
\section{Summary}\label{sec:summary}
%======================================================================
\subsection{Conclusion}\label{subsec:conclusions}

This work shows that \emph{\HCFfull\ (HCF) code}—%
specified by \weighttwo measurements executed in a \sixstep{}
periodic schedule on \threecolorable \(\{p,3\}\) tilings—constitutes Floquet code with \finiteenc. 
Because each single-qubit fault excites at most two detectors, the resulting
\graphedgesyn{} can be decoded by \MWPM{} in
a near linear way in $n$ (empirically $\approx O(n)$) time, thus avoiding the \hyperedgesyn{}
overhead that slows conventional HF code decoding.
The \sixstep{} preserves the underlying anyon structure and
suppresses high-degree detector events, leading to an estimated circuit-level
threshold \(\approx\! 1.5\,\%\).
Semi-hyperbolic refinement can increase the code distance by an integer
factor \(\ell\) without increasing the check weight.
Under the \EMthree{} noise model, \HCF\ code achieves a lower
\logicalrate{} than \HF\ code of identical size.

%-----------------------------------------------------------------------
\subsection{Outlook}\label{subsec:outlook}
%-----------------------------------------------------------------------

One potential direction for future work is the development of a planar adaptation of the HCF code, i.e., realizing boundary-terminated patches.
Boundary terminations and schedule designs for planar Floquet codes have already been explored~\cite{Vuillot2021Planar,Gidney2022Planar,Haah2022Boundaries}.
Building on these ideas, it appears feasible that a planar HCF code could enable lattice surgery and twist defects while retaining weight-2 checks.
Such a construction could support compact on-chip patch layouts and extensions to patch-to-patch lattice surgery~\cite{Sutcliffe2025DistributedQEC}, potentially facilitating inter-module entanglement links.

An additional direction for future work is to extend the noise modeling applied to the HCF code beyond the EM3 model assumptions.
In practical devices, spatial correlations (e.g., crosstalk), temporal correlations (e.g., non-Markovian), and non-Pauli channels (e.g., erasure, amplitude damping, and leakage) may not be negligible.
Introducing a parametric family of circuit-level models with EM3 as a special case could enable a controlled assessment of how these effects influence thresholds and decoding latency.
On the decoder side, it may be possible to retain an MWPM-based workflow by correlation-aware reweighting of spatio-temporal edges. Where necessary, this could be augmented with lightweight BP initialization~\cite{Higgott2023BeliefMatching} or low-order OSD to accommodate more complex detector graphs.
In parallel, recent analytical work on competing automorphisms in disordered Floquet codes~\cite{AitchisonBeri2025} may provide a complementary framework for reasoning about measurement imperfections and correlated faults in schedules of the HCF code.

Exploring routes to non-Clifford operations within the HCF framework presents another significant challenge.
Recent dynamical, measurement-based approaches offer possible templates. These include: (i) a rewinding-based construction that realizes a transversal logical CCZ in stacked 3D Floquet codes~\cite{Dua2024Rewinding}; (ii) Dynamic Automorphism (DA) codes that implement non-Clifford gates via adaptive low-weight measurements in their 3D variants~\cite{Davydova2024Dynamic}; and (iii) protocols within strictly 2D architectures that use domain walls between the surface code and a non-Abelian phase to implement non-Clifford gates (including CCZ and $T$), together with an associated threshold theorem~\cite{Davydova2025WithoutTiedKnots}.
It remains an open question whether these ideas can be adapted to the HCF code while maintaining compatibility with the HCF structure. Nevertheless, such an adaptation could, in principle, provide a route toward universal quantum computation.

Finally, demonstrations of Floquet codes on superconducting hardware have been reported, 
providing early evidence that periodic schedules of weight-2 checks are experimentally 
achievable~\cite{Wootton2022Floquet,Sun2025FloquetBaconShor}.
As plausible next steps, it would be valuable to explore alternative measurement 
schedules on these platforms and to extend these investigations to other quantum 
computing architectures—including trapped-ion, photonic, and neutral-atom systems, 
among others—to characterize platform-dependent processes for Floquet QEC. 

%======================================================================
\section*{Data availability}
%======================================================================
The data and simulation code are available from the corresponding
author upon reasonable request.

%======================================================================
\section*{Acknowledgments}
%======================================================================
The authors appreciate their colleagues at the institute for many insightful discussions and for their continuous support throughout this work.

\bibliographystyle{quantum}
\bibliography{references}

\begin{thebibliography}{10}

\bibitem{Shor1997}
Peter~W. Shor.
\newblock ``Polynomial-time algorithms for prime factorization and discrete logarithms on a quantum computer''.
\newblock \href{https://dx.doi.org/10.1137/S0097539795293172}{SIAM Journal on Computing {\bf 26}, 1484--1509}~(1997).

\bibitem{Georgescu2014}
I.~M. Georgescu, S.~Ashhab, and F.~Nori.
\newblock ``Quantum simulation''.
\newblock \href{https://dx.doi.org/10.1103/RevModPhys.86.153}{Reviews of Modern Physics {\bf 86}, 153--185}~(2014).
\newblock  \href{http://arxiv.org/abs/1308.6253}{arXiv:1308.6253}.

\bibitem{Gharibian2022}
Sevag Gharibian and Fran{\c{c}}ois~Le Gall.
\newblock ``Dequantizing the quantum singular value transformation: hardness and applications to quantum chemistry and the quantum pcp conjecture''.
\newblock In Proceedings of the 54th Annual ACM SIGACT Symposium on Theory of Computing (STOC 2022).
\newblock \href{https://dx.doi.org/10.1145/3519935.3519991}{Pages 19--32}.
\newblock Rome, Italy~(2022). ACM.

\bibitem{Harrow2009}
Aram~W. Harrow, Avinatan Hassidim, and Seth Lloyd.
\newblock ``Quantum algorithm for linear systems of equations''.
\newblock \href{https://dx.doi.org/10.1103/PhysRevLett.103.150502}{Physical Review Letters {\bf 103}, 150502}~(2009).
\newblock  \href{http://arxiv.org/abs/0811.3171}{arXiv:0811.3171}.

\bibitem{Cai2023}
Z.~Cai, R.~Babbush, S.~C. Benjamin, S.~Endo, W.~J. Huggins, Y.~Li, J.~R. McClean, and T.~E. O'Brien.
\newblock ``Quantum error mitigation''.
\newblock \href{https://dx.doi.org/10.1103/RevModPhys.95.045005}{Reviews of Modern Physics {\bf 95}, 045005}~(2023).
\newblock  \href{http://arxiv.org/abs/2210.00921}{arXiv:2210.00921}.

\bibitem{Nielsen2010}
Michael~A. Nielsen and Isaac~L. Chuang.
\newblock ``Quantum computation and quantum information: 10th anniversary edition''.
\newblock \href{https://dx.doi.org/10.1017/CBO9780511976667}{Cambridge University Press}. Cambridge~(2010).

\bibitem{Preskill2018}
John Preskill.
\newblock ``Quantum computing in the {NISQ} era and beyond''.
\newblock \href{https://dx.doi.org/10.22331/q-2018-08-06-79}{Quantum {\bf 2}, 79}~(2018).
\newblock  \href{http://arxiv.org/abs/1801.00862}{arXiv:1801.00862}.

\bibitem{Takagi2022}
Ryuji Takagi, Suguru Endo, Shintaro Minagawa, and Mile Gu.
\newblock ``Fundamental limits of quantum error mitigation''.
\newblock \href{https://dx.doi.org/10.1038/s41534-022-00618-z}{npj Quantum Information {\bf 8}, 114}~(2022).

\bibitem{Tsubouchi2023}
Kento Tsubouchi, Takahiro Sagawa, and Naoki Yoshioka.
\newblock ``Universal cost bound of quantum error mitigation based on quantum estimation theory''.
\newblock \href{https://dx.doi.org/10.1103/PhysRevLett.131.210601}{Physical Review Letters {\bf 131}, 210601}~(2023).

\bibitem{Takagi2023}
Ryuji Takagi, Hiroyasu Tajima, and Mile Gu.
\newblock ``Universal sampling lower bounds for quantum error mitigation''.
\newblock \href{https://dx.doi.org/10.1103/PhysRevLett.131.210602}{Physical Review Letters {\bf 131}, 210602}~(2023).

\bibitem{Quek2024}
Yihui Quek, Daniel~Stilck Fran{\c{c}}a, Sumeet Khatri, Johannes~Jakob Meyer, and Jens Eisert.
\newblock ``Exponentially tighter bounds on limitations of quantum error mitigation''.
\newblock \href{https://dx.doi.org/10.1038/s41567-024-02536-7}{Nature Physics {\bf 20}, 1648--1658}~(2024).

\bibitem{Suzuki2022}
Yasunari Suzuki, Suguru Endo, Keisuke Fujii, and Yuuki Tokunaga.
\newblock ``Quantum error mitigation as a universal error reduction technique: Applications from the nisq to the fault-tolerant quantum computing era''.
\newblock \href{https://dx.doi.org/10.1103/PRXQuantum.3.010345}{PRX Quantum {\bf 3}, 010345}~(2022).

\bibitem{Gottesman1997Stabilizer}
Daniel Gottesman.
\newblock ``{Stabilizer Codes and Quantum Error Correction}''.
\newblock \href{https://dx.doi.org/10.48550/ARXIV.quant-ph/9705052}{PhD thesis}.
\newblock California Institute of Technology.
\newblock ~(1997).
\newblock  \href{http://arxiv.org/abs/quant-ph/9705052}{arXiv:quant-ph/9705052}.

\bibitem{Knill1997Theory}
Emanuel Knill and Raymond Laflamme.
\newblock ``{Theory of Quantum Error-Correcting Codes}''.
\newblock \href{https://dx.doi.org/10.1103/PhysRevA.55.900}{Physical Review A {\bf 55}, 900--911}~(1997).
\newblock  \href{http://arxiv.org/abs/quant-ph/9604034}{arXiv:quant-ph/9604034}.

\bibitem{Calderbank1996Good}
A.~R. Calderbank and Peter~W. Shor.
\newblock ``{Good Quantum Error-Correcting Codes Exist}''.
\newblock \href{https://dx.doi.org/10.1103/PhysRevA.54.1098}{Physical Review A {\bf 54}, 1098--1105}~(1996).
\newblock  \href{http://arxiv.org/abs/quant-ph/9512032}{arXiv:quant-ph/9512032}.

\bibitem{Dennis2002Topological}
Eric Dennis, Alexei Kitaev, Andrew Landahl, and John Preskill.
\newblock ``{Topological Quantum Memory}''.
\newblock \href{https://dx.doi.org/10.1063/1.1499754}{Journal of Mathematical Physics {\bf 43}, 4452--4505}~(2002).
\newblock  \href{http://arxiv.org/abs/quant-ph/0110143}{arXiv:quant-ph/0110143}.

\bibitem{Bombin2006}
H\'{e}ctor Bomb\'{\i}n and Miguel~A. Mart\'{\i}n-Delgado.
\newblock ``Topological quantum distillation''.
\newblock \href{https://dx.doi.org/10.1103/PhysRevLett.97.180501}{Physical Review Letters {\bf 97}, 180501}~(2006).
\newblock  \href{http://arxiv.org/abs/quant-ph/0}{arXiv:quant-ph/060513}.

\bibitem{Panteleev2021Goodcode}
Pavel Panteleev and Gleb Kalachev.
\newblock ``Asymptotically good quantum and locally testable classical ldpc codes''.
\newblock In Proceedings of the 54th Annual ACM SIGACT Symposium on Theory of Computing (STOC 2022).
\newblock \href{https://dx.doi.org/10.1145/3519935.3520017}{Pages 375--388}.
\newblock ~(2022).
\newblock  \href{http://arxiv.org/abs/2111.03654}{arXiv:2111.03654}.

\bibitem{Bacon2006Operator}
Dave Bacon.
\newblock ``{Operator Quantum Error-Correcting Subsystems for Self-Correcting Quantum Memories}''.
\newblock \href{https://dx.doi.org/10.1103/PhysRevA.73.012340}{Physical Review A {\bf 73}, 012340}~(2006).
\newblock  \href{http://arxiv.org/abs/quant-ph/0506023}{arXiv:quant-ph/0506023}.

\bibitem{Paetznick2013Universal}
Adam Paetznick and Ben~W. Reichardt.
\newblock ``Universal fault-tolerant quantum computation with only transversal gates and error correction''.
\newblock \href{https://dx.doi.org/10.1103/PhysRevLett.111.090505}{Physical Review Letters {\bf 111}, 090505}~(2013).
\newblock  \href{http://arxiv.org/abs/1304.3709}{arXiv:1304.3709}.

\bibitem{Chamberland2020Topological}
Christopher Chamberland, Guanyu Zhu, Theodore~J. Yoder, Jared~B. Hertzberg, and Andrew~W. Cross.
\newblock ``{Topological and Subsystem Codes on Low-Degree Graphs with Flag Qubits}''.
\newblock \href{https://dx.doi.org/10.1103/PhysRevX.10.011022}{Physical Review X {\bf 10}, 011022}~(2020).
\newblock  \href{http://arxiv.org/abs/1907.09528}{arXiv:1907.09528}.

\bibitem{Aharonov2013}
Dorit Aharonov, Itai Arad, and Thomas Vidick.
\newblock ``Guest column: The quantum pcp conjecture''.
\newblock \href{https://dx.doi.org/10.1145/2491533.2491549}{ACM SIGACT News {\bf 44}, 47--79}~(2013).

\bibitem{Anshu2023}
Anurag Anshu, Nikolas~P. Breuckmann, and Chinmay Nirkhe.
\newblock ``Nlts hamiltonians from good quantum codes''.
\newblock In Proceedings of the 55th Annual ACM Symposium on Theory of Computing (STOC 2023).
\newblock \href{https://dx.doi.org/10.1145/3564246.3585114}{Pages 1090--1096}.
\newblock Orlando, FL, USA~(2023). ACM.

\bibitem{Krinner2022ErrorBudget}
Sebastian Krinner, Nathan Lacroix, Ants Remm, Agustin~Di Paolo, Elie Genois, Catherine Leroux, Christoph Hellings, Stefania Lazar, Francois Swiadek, Johannes Herrmann, Graham~J. Norris, Christian~Kraglund Andersen, Markus M{\"u}ller, Alexandre Blais, Christopher Eichler, and Andreas Wallraff.
\newblock ``Realizing repeated quantum error correction in a distance-three surface code''.
\newblock \href{https://dx.doi.org/10.1038/s41586-022-04566-8}{Nature {\bf 605}, 669--674}~(2022).

\bibitem{Arute2019Supremacy}
Frank Arute, Kunal Arya, Ryan Babbush, Dave Bacon, Joseph~C. Bardin, Rami Barends, Rupak Biswas, Sergio Boixo, Fernando G. S.~L. Brandao, David~A. Buell, et~al.
\newblock ``Quantum supremacy using a programmable superconducting processor''.
\newblock \href{https://dx.doi.org/10.1038/s41586-019-1666-5}{Nature {\bf 574}, 505--510}~(2019).

\bibitem{Wright2019Benchmarking}
K.~Wright, K.~M. Beck, S.~Debnath, J.~M. Amini, Y.~Nam, N.~Grzesiak, J.-S. Chen, N.~C. Pisenti, M.~Chmielewski, C.~Collins, et~al.
\newblock ``Benchmarking an 11-qubit quantum computer''.
\newblock \href{https://dx.doi.org/10.1038/s41467-019-13534-2}{Nature Communications {\bf 10}, 5464}~(2019).

\bibitem{RyanAnderson2021RealTime}
C.~Ryan-Anderson, J.~G. Bohnet, K.~Lee, D.~Gresh, A.~Hankin, J.~P. Gaebler, D.~Francois, A.~Chernoguzov, D.~Lucchetti, N.~C. Brown, et~al.
\newblock ``Realization of real-time fault-tolerant quantum error correction''.
\newblock \href{https://dx.doi.org/10.1103/PhysRevX.11.041058}{Physical Review X {\bf 11}, 041058}~(2021).

\bibitem{Egan2021FaultTolerant}
Laird Egan, Dripto~M. Debroy, Crystal Noel, Andrew Risinger, Daiwei Zhu, Debopriyo Biswas, Michael Newman, Muyuan Li, Kenneth~R. Brown, Marko Cetina, and Christopher Monroe.
\newblock ``Fault-tolerant control of an error-corrected qubit''.
\newblock \href{https://dx.doi.org/10.1038/s41586-021-03928-y}{Nature {\bf 598}, 281--286}~(2021).

\bibitem{HastingsHaah2021Dynamical}
Matthew~B. Hastings and Jeongwan Haah.
\newblock ``Dynamically generated logical qubits''.
\newblock \href{https://dx.doi.org/10.22331/q-2021-11-02-564}{Quantum {\bf 5}, 564}~(2021).
\newblock  \href{http://arxiv.org/abs/2107.02194}{arXiv:2107.02194}.

\bibitem{Vuillot2021Planar}
Christophe Vuillot.
\newblock ``Planar floquet codes''~(2021).
\newblock  \href{http://arxiv.org/abs/2110.05348}{arXiv:2110.05348}.

\bibitem{Gidney2022Planar}
Craig Gidney, Michael Newman, and Matt McEwen.
\newblock ``Benchmarking the planar honeycomb code''.
\newblock \href{https://dx.doi.org/10.22331/q-2022-09-21-813}{Quantum {\bf 6}, 813}~(2022).
\newblock  \href{http://arxiv.org/abs/2202.11845}{arXiv:2202.11845}.

\bibitem{Haah2022Boundaries}
Jeongwan Haah and Matthew~B. Hastings.
\newblock ``Boundaries for the honeycomb code''.
\newblock \href{https://dx.doi.org/10.22331/q-2022-04-21-693}{Quantum {\bf 6}, 693}~(2022).
\newblock  \href{http://arxiv.org/abs/2110.09545}{arXiv:2110.09545}.

\bibitem{Davydova2023}
Margarita Davydova, Nathanan Tantivasadakarn, and Shankar Balasubramanian.
\newblock ``Floquet codes without parent subsystem codes''.
\newblock \href{https://dx.doi.org/10.1103/PRXQuantum.4.020341}{PRX Quantum {\bf 4}, 020341}~(2023).
\newblock  \href{http://arxiv.org/abs/2210.02468}{arXiv:2210.02468}.

\bibitem{Kesselring2024}
Markus~S. Kesselring, Julio~C. Magdalena de~la Fuente, Felix Thomsen, Jens Eisert, Stephen~D. Bartlett, and Benjamin~J. Brown.
\newblock ``Anyon condensation and the color code''.
\newblock \href{https://dx.doi.org/10.1103/PRXQuantum.5.010342}{PRX Quantum {\bf 5}, 010342}~(2024).
\newblock  \href{http://arxiv.org/abs/2212.00042}{arXiv:2212.00042}.

\bibitem{Higgott2024SemiHyper}
Oscar Higgott and Nikolas~P. Breuckmann.
\newblock ``Constructions and performance of hyperbolic and semi-hyperbolic floquet codes''.
\newblock \href{https://dx.doi.org/10.1103/PRXQuantum.5.040327}{PRX Quantum {\bf 5}, 040327}~(2024).
\newblock  \href{http://arxiv.org/abs/2308.03750}{arXiv:2308.03750}.

\bibitem{Fahimniya2024}
Ali Fahimniya, Hossein Dehghani, Kishor Bharti, Sheryl Mathew, Alicia~J. Koll\'ar, Alexey~V. Gorshkov, and Michael~J. Gullans.
\newblock ``Fault-tolerant hyperbolic floquet quantum error correcting codes''.
\newblock \href{https://dx.doi.org/10.1103/PhysRevLett.132.091903}{Physical Review Letters {\bf 132}, 091903}~(2024).
\newblock  \href{http://arxiv.org/abs/2309.10033}{arXiv:2309.10033}.

\bibitem{Higgott2025sparseblossom}
Oscar Higgott and Craig Gidney.
\newblock ``Sparse blossom: correcting a million errors per core second with minimum-weight matching''.
\newblock \href{https://dx.doi.org/10.22331/q-2025-01-20-1600}{Quantum {\bf 9}, 1600}~(2025).
\newblock  \href{http://arxiv.org/abs/2303.15933}{arXiv:2303.15933}.

\bibitem{Gottesman1998Theory}
Daniel Gottesman.
\newblock ``{Theory of Fault-Tolerant Quantum Computation}''.
\newblock \href{https://dx.doi.org/10.1103/PhysRevA.57.127}{Physical Review A {\bf 57}, 127--137}~(1998).
\newblock  \href{http://arxiv.org/abs/quant-ph/9702029}{arXiv:quant-ph/9702029}.

\bibitem{Steane1996Error}
Andrew~M. Steane.
\newblock ``{Error Correcting Codes in Quantum Theory}''.
\newblock \href{https://dx.doi.org/10.1103/PhysRevLett.77.793}{Physical Review Letters {\bf 77}, 793--797}~(1996).

\bibitem{Poulin2005Stabilizer}
David Poulin.
\newblock ``{Stabilizer Formalism for Operator Quantum Error Correction}''.
\newblock \href{https://dx.doi.org/10.1103/PhysRevLett.95.230504}{Physical Review Letters {\bf 95}, 230504}~(2005).
\newblock  \href{http://arxiv.org/abs/quant-ph/0508131}{arXiv:quant-ph/0508131}.

\bibitem{Bombin2015Gauge}
H{\'e}ctor Bomb{\'\i}n.
\newblock ``{Gauge Color Codes: Optimal Transversal Gates and Gauge Fixing in Topological Stabilizer Codes}''.
\newblock \href{https://dx.doi.org/10.1088/1367-2630/17/8/083002}{New Journal of Physics {\bf 17}, 083002}~(2015).
\newblock  \href{http://arxiv.org/abs/1311.0879}{arXiv:1311.0879}.

\bibitem{Brown2016Fault}
Benjamin~J. Brown, Naomi~H. Nickerson, and Dan~E. Browne.
\newblock ``{Fault-Tolerant Error Correction with the Gauge Color Code}''.
\newblock \href{https://dx.doi.org/10.1038/ncomms12302}{Nature Communications {\bf 7}, 12302}~(2016).
\newblock  \href{http://arxiv.org/abs/1503.08217}{arXiv:1503.08217}.

\bibitem{FuGottesman2024ECinDynamicalCodes}
Xiaozhen Fu and Daniel Gottesman.
\newblock ``Error correction in dynamical codes''~(2024).
\newblock  \href{http://arxiv.org/abs/2403.04163}{arXiv:2403.04163}.

\bibitem{Gidney2021Memory}
Craig Gidney, Michael Newman, Austin Fowler, and Michael Broughton.
\newblock ``A fault-tolerant honeycomb memory''.
\newblock \href{https://dx.doi.org/10.22331/q-2021-12-20-605}{Quantum {\bf 5}, 605}~(2021).
\newblock  \href{http://arxiv.org/abs/2108.10457}{arXiv:2108.10457}.

\bibitem{setiawan2024}
F.~Setiawan and Campbell McLauchlan.
\newblock ``Tailoring dynamical codes for biased noise: The $x^3z^3$ floquet code''~(2024).
\newblock  \href{http://arxiv.org/abs/2411.04974}{arXiv:2411.04974}.

\bibitem{Breuckmann2017Hyperbolic}
Nikolas~P. Breuckmann, Christophe Vuillot, Earl~T. Campbell, Anirudh Krishna, and Barbara~M. Terhal.
\newblock ``Hyperbolic and semi-hyperbolic surface codes for quantum storage''.
\newblock \href{https://dx.doi.org/10.1088/2058-9565/aa7d3b}{Quantum Science and Technology {\bf 2}, 035007}~(2017).
\newblock  \href{http://arxiv.org/abs/1703.00590}{arXiv:1703.00590}.

\bibitem{Kitaev2003Anyons}
Alexei~Yu. Kitaev.
\newblock ``Fault-tolerant quantum computation by anyons''.
\newblock \href{https://dx.doi.org/10.1016/S0003-4916(02)00018-0}{Annals of Physics {\bf 303}, 2--30}~(2003).
\newblock  \href{http://arxiv.org/abs/quant-ph/9}{arXiv:quant-ph/970702}.

\bibitem{Kesselring2018ColorBoundaries}
Markus~S. Kesselring, Fernando Pastawski, Jens Eisert, and Benjamin~J. Brown.
\newblock ``The boundaries and twist defects of the color code and their applications to topological quantum computation''.
\newblock \href{https://dx.doi.org/10.22331/q-2018-10-19-101}{Quantum {\bf 2}, 101}~(2018).
\newblock  \href{http://arxiv.org/abs/1806.02820}{arXiv:1806.02820}.

\bibitem{Fowler2012Surface}
Austin~G. Fowler, Matteo Mariantoni, John~M. Martinis, and Andrew~N. Cleland.
\newblock ``{Surface Codes: Towards Practical Large-Scale Quantum Computation}''.
\newblock \href{https://dx.doi.org/10.1103/PhysRevA.86.032324}{Physical Review A {\bf 86}, 032324}~(2012).
\newblock  \href{http://arxiv.org/abs/1208.0928}{arXiv:1208.0928}.

\bibitem{Panteleev2021}
Pavel Panteleev and Gleb Kalachev.
\newblock ``Degenerate quantum ldpc codes with good finite length performance''.
\newblock \href{https://dx.doi.org/10.22331/q-2021-11-22-585}{Quantum {\bf 5}, 585}~(2021).
\newblock  \href{http://arxiv.org/abs/1904.02703}{arXiv:1904.02703}.

\bibitem{roffe2020BPOSD}
Joschka Roffe, David~R. White, Simon Burton, and Earl~T. Campbell.
\newblock ``Decoding across the quantum low-density parity-check code landscape''.
\newblock \href{https://dx.doi.org/10.1103/PhysRevResearch.2.043423}{Physical Review Research {\bf 2}, 043423}~(2020).

\bibitem{DerksEtAl2025NoisyReadouts}
Peter-Jan H.~S. Derks, Alex Townsend-Teague, Jens Eisert, Markus~S. Kesselring, Oscar Higgott, and Benjamin~J. Brown.
\newblock ``Dynamical codes for hardware with noisy readouts''~(2025).
\newblock  \href{http://arxiv.org/abs/2505.07658}{arXiv:2505.07658}.

\bibitem{Sutcliffe2025DistributedQEC}
Evan Sutcliffe, Bhargavi Jonnadula, Claire Le~Gall, Alexandra~E. Moylett, and Coral~M. Westoby.
\newblock ``Distributed quantum error correction based on hyperbolic floquet codes''~(2025).
\newblock  \href{http://arxiv.org/abs/2501.14029}{arXiv:2501.14029}.

\bibitem{Higgott2023BeliefMatching}
Oscar Higgott, Thomas~C. Bohdanowicz, Aleksander Kubica, Steven~T. Flammia, and Earl~T. Campbell.
\newblock ``Improved decoding of circuit noise and fragile boundaries of tailored surface codes''.
\newblock \href{https://dx.doi.org/10.1103/PhysRevX.13.031007}{Physical Review X {\bf 13}, 031007}~(2023).
\newblock  \href{http://arxiv.org/abs/2203.04948}{arXiv:2203.04948}.

\bibitem{AitchisonBeri2025}
Cory~T. Aitchison and Benjamin B{\'e}ri.
\newblock ``Competing automorphisms and disordered floquet codes''.
\newblock \href{https://dx.doi.org/10.1103/PhysRevB.111.235112}{Physical Review B {\bf 111}, 235112}~(2025).

\bibitem{Dua2024Rewinding}
Arpit Dua, Nathanan Tantivasadakarn, Joseph Sullivan, and Tyler~D. Ellison.
\newblock ``Engineering 3d floquet codes by rewinding''.
\newblock \href{https://dx.doi.org/10.1103/PRXQuantum.5.020305}{PRX Quantum {\bf 5}, 020305}~(2024).
\newblock  \href{http://arxiv.org/abs/2307.13668}{arXiv:2307.13668}.

\bibitem{Davydova2024Dynamic}
Margarita Davydova, Nathanan Tantivasadakarn, Shankar Balasubramanian, and David Aasen.
\newblock ``Quantum computation from dynamic automorphism codes''.
\newblock \href{https://dx.doi.org/10.22331/q-2024-08-27-1448}{Quantum {\bf 8}, 1448}~(2024).
\newblock  \href{http://arxiv.org/abs/2307.10353}{arXiv:2307.10353}.

\bibitem{Davydova2025WithoutTiedKnots}
Margarita Davydova, Andreas Bauer, Julio C.~Magdalena de~la Fuente, Mark Webster, Dominic~J. Williamson, and Benjamin~J. Brown.
\newblock ``Universal fault tolerant quantum computation in 2d without getting tied in knots''~(2025).
\newblock  \href{http://arxiv.org/abs/2503.15751}{arXiv:2503.15751}.

\bibitem{Wootton2022Floquet}
James~R. Wootton.
\newblock ``Measurements of floquet code plaquette stabilizers''~(2022) \href{http://arxiv.org/abs/2210.13154}{arXiv:2210.13154}.

\bibitem{Sun2025FloquetBaconShor}
Xuandong Sun et~al.
\newblock ``Logical operations with a dynamical qubit in floquet-bacon-shor code''~(2025).
\newblock  \href{http://arxiv.org/abs/2503.03867}{arXiv:2503.03867}.

\end{thebibliography}

\clearpage 
\onecolumn    

\appendix  

%========================
\section{Detector weight}
\label{app:detector_weight}
%========================

\subsection{Detector weight of \HF\ code}

\begin{figure}[H]
  \centering
  \includegraphics[width=0.7\linewidth]{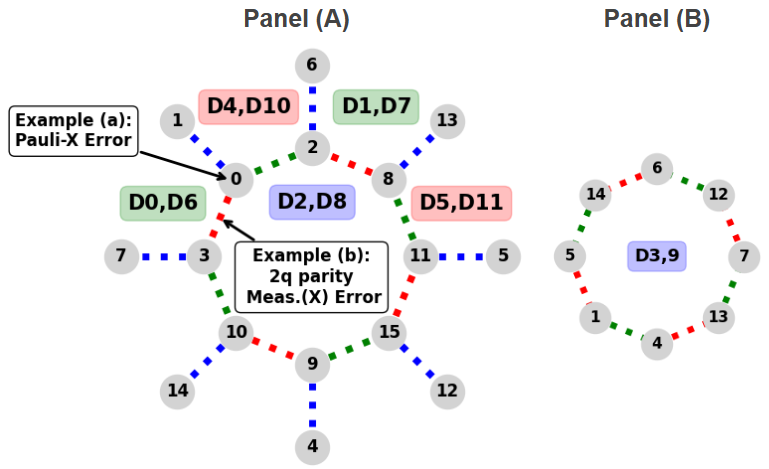}
\caption{The definition of an \HF{} detector on a 16-qubit lattice.
Panel (A) shows the lattice in the orientation where red/green \plaquette s face the viewer; panel (B) depicts the “back side” of (A).
Gray nodes are the physical qubits (0–15).
Edge colors indicate the measured Pauli in \HF{} code, with
\textcolor{red!70!black}{red}=\(\mathrm{XX}\) (rXX),
\textcolor{green!60!black}{green}=\(\mathrm{YY}\) (gYY),
and \textcolor{blue!70!black}{blue}=\(\mathrm{ZZ}\) (bZZ)
(edges drawn as dotted lines).
Labels with colored backgrounds denote detector groups:
\textcolor{blue!70!black}{blue}=\{D2, D8 ,D3, D9\},
\textcolor{green!60!black}{green}=\{D0, D6, D1, D7\},
\textcolor{red!70!black}{red}=\{D4, D10, D5, D11\}.}
  \label{fig:hf-detector-map}
\end{figure}

At \HF{} code, detectors are defined as the differences of consecutive checks under the schedule $\mathrm{rXX}\!\to\!\mathrm{gYY}\!\to\!\mathrm{bZZ}$.
We illustrate how the type of error affects the detector weight.
When temporal boundaries truncate the window, odd weights can appear.
Figure~\ref{fig:hf-detector-map} and the examples below summarize these patterns.

% -------------------------------------------

\begin{figure}[H]
\centering
\includegraphics[width=0.92\linewidth]{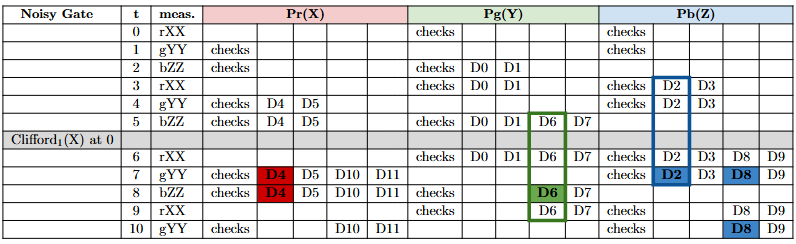}
\caption{Example (a): A data \qubit error on \HF{} code.
The gray band indicates the error 
insertion interval between $t=5$ and $t=6$, where a Pauli-X error occurs 
on qubit 0. Colored backgrounds (green for D6, blue for D2) highlight 
the checks triggered by this single-error event. Colored frames delineate 
the detector ranges containing these checks. A single data qubit error 
produces two simultaneous, spatially adjacent detector triggers.}
\label{fig:hf-data}
\end{figure}

\paragraph{Example (a): Data \qubit error in \HF\ code}
A data qubit error manifests as a persistent state change that remains until correction. Consider a single data error occur at qubit 0. 
Figure~\ref{fig:hf-detector-map} illustrates the spatial configuration of detector responses to a single $X$ error applied to qubit 0 during the interval $t \in (5,6)$. 
The gray band in Figure~\ref{fig:hf-data} indicates the temporal location of this error at time step $t \in (5,6)$, showing how the error propagates through successive measurement steps. 
This error anti-commutes with Z-type stabilizer checks while commuting with X-type checks, resulting in a $w=2$ detection signature.

Under the measurement schedule $\mathcal{M}_1(\mathrm{XX}) \to 
\mathcal{M}_2(\mathrm{YY}) \to \mathcal{M}_3(\mathrm{ZZ})$, a single
$X$ error occurring at time $t = n$ affects the detector parities at two
consecutive time steps. Specifically, it triggers detectors at 
$\mathcal{M}_2(\mathrm{YY})$ (time $t = n+1$) and $\mathcal{M}_3(\mathrm{ZZ})$ 
(time $t = n+2$), corresponding to the anti-commuting stabilizers.

\begin{figure}[H]
\centering
\includegraphics[width=0.92\linewidth]{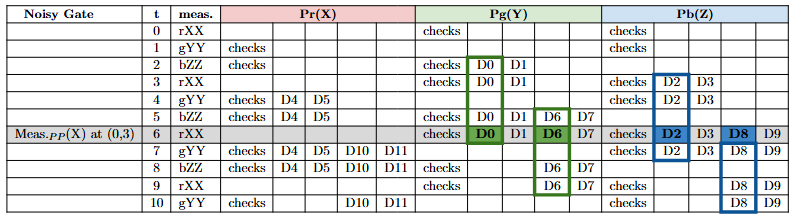}
\caption{Example (b): Measurement error in the HF code.
The gray band indicates time step $t=6$ where a parity measurement error occurs at qubits 0 and 3. Colored backgrounds highlight 
triggered checks: D0 and D6 (green) and D2 and D8 (blue). Colored frames 
delineate the detector ranges containing these checks. Because detectors 
are defined as differences between consecutive measurements, triggers 
appear both when the detector value is recorded and when the correct 
value is restored, resulting in a total weight of $w=4$ for this single 
measurement error.}
\label{fig:HF_MPP}
\end{figure}

\paragraph{Example (b): Measurement error in the HF code}
A measurement error, which arises from error two-body parity measurements on data qubits, is transient—affecting only the specific measurement step in which it occurs. This contrasts with a data qubit error, which persists until correction.
Consider a single measurement error affecting the XX parity check at time $t=6$. 
Figure~\ref{fig:hf-detector-map} illustrates a $X$ measurement error applied to qubits 0 and 3. 
The gray band (see Figure~\ref{fig:HF_MPP}) indicates the time step $t=6$ where a parity measurement error occurs at qubits 0 and 3. 

This error is immediately incorporated into the syndrome extraction, causing discrepancies with the syndrome values from adjacent time steps. Since detectors are triggered by changes between consecutive stabilizer measurements, this single measurement error generates four detector activations.

\paragraph{Boundary cases: $w=1$ and $3$.}
Because detectors are defined as differences between consecutive checks, 
temporal boundaries can truncate even-weight patterns and produce odd weights.
For $w=1$, a data qubit error in the first or last time step yields only one
detector trigger, appearing as an edge to a temporal boundary node in the
detector graph. For $w=3$, the temporal truncation of the boundary errors reduces
the nominal $w=4$ triggers by one.

% -------------------------------------------

% -------------------------------------------

\begin{figure}[H]
\centering
\includegraphics[width=0.92\linewidth]{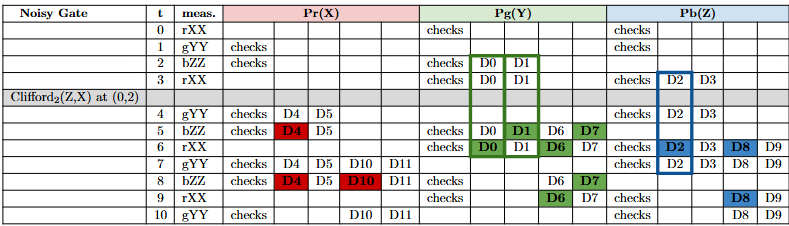}
\caption{Boundary-induced odd weight ($w=3$) from a correlated data qubit 
error. In the HF schedule $\mathrm{rXX} \to \mathrm{gYY} \to \mathrm{bZZ}$, 
a two-qubit error $\mathrm{Clifford}_2(Z,X)$ on qubits (0,2) is injected 
at $t \in (3,4)$. The $Z_0$ component triggers detectors $D_2$ at $t=6$ 
and $D_0$ at $t=6$. The $X_2$ component triggers $D_1$ at $t=5$.}
\label{fig:HF_w3}
\end{figure}

\clearpage

% -------------------------------------------
\subsection{Detector weights in HCF code}
\label{sec:css-detector-weight}
\begin{figure}[H]
\centering
\includegraphics[width=0.7\linewidth]{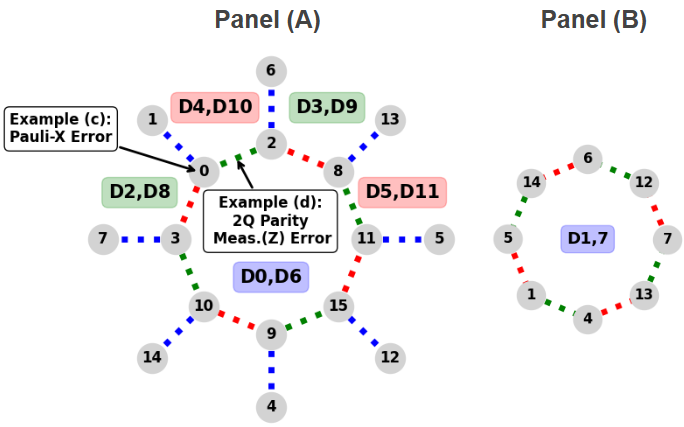}
\caption{\textbf{Pauli-Z detector structure of HCF code.}
Gray nodes represent physical qubits (labeled 0--15). Dotted edges indicate the measured Pauli operators. The measurement schedule alternates between $\{\mathrm{XX}, \mathrm{ZZ}\}$ operations at each time step, independent of the detector's Pauli type (X or Z).
The color coding represents the temporal scheduling:
\textcolor{red!70!black}{red} = $\mathrm{XX}/\mathrm{ZZ}$,
\textcolor{green!60!black}{green} = $\mathrm{XX}/\mathrm{ZZ}$,
\textcolor{blue!70!black}{blue} = $\mathrm{XX}/\mathrm{ZZ}$.
Colored labels denote Pauli-Z detector groups according to the measurement schedule:
\textcolor{blue!70!black}{blue} = $\{\mathrm{D}_0, \mathrm{D}_6, \mathrm{D}_1, \mathrm{D}_7\}$,
\textcolor{green!60!black}{green} = $\{\mathrm{D}_2, \mathrm{D}_8, \mathrm{D}_3, \mathrm{D}_9\}$,
\textcolor{red!70!black}{red} = $\{\mathrm{D}_4, \mathrm{D}_{10}, \mathrm{D}_5, \mathrm{D}_{11}\}$.}
\label{fig:hcf-z-detectors}
\end{figure}

This section examines the detector-weight properties of HCF code using 
time-gapped detector construction, as illustrated in 
Figure~\ref{fig:hcf-z-detectors}. The analysis focuses on Z-type 
detectors for clarity.

\paragraph{Example (c): Data qubit error in HCF code}
A data qubit error represents a persistent state modification that persists until corrected. Consider a single data error occur at qubit 0.
Figure~\ref{fig:hcf-z-detectors} displays the spatial pattern of detector triggers from a single $X$ error on qubit 0 during the interval $t \in (6,7)$. 
The error creates temporal boundaries at the affected Z-type plaquettes, 
triggering two detectors. Detector D0 corresponds to a gapped Z-check 
spanning the rZZ measurement at $t=3$ and the gZZ measurement at $t=7$. 
Detector D2 activates from the boundary between bZZ at $t=5$ and rZZ measurements 
at $t=9$. This pair yields the minimal detector weight $w=2$ for 
non-trivial errors in this code.

\begin{figure}[H]
\centering
\includegraphics[width=0.85\linewidth]{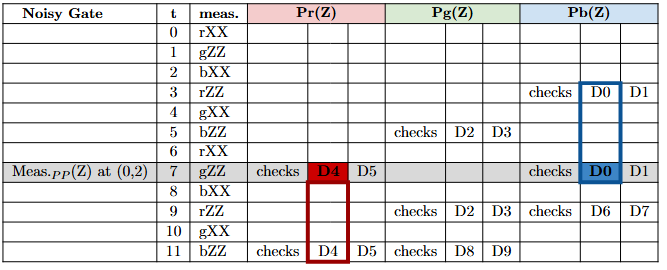}
\caption{Example (d): Measurement error in HCF code.
The gray band indicates time 
step $t=7$ where the measurement error occurs. Colored backgrounds 
indicate triggered detectors: D4 (red) and D0 (blue), with colored 
frames showing the detector ranges. Under time-gapped detector 
construction, this single measurement error triggers two 
detectors, yielding $w=2$ in this noise model.}
\label{fig:hcf-meas}
\end{figure}

\paragraph{Example (d): Measurement error in HCF code}
In HCF code, consider a single measurement error affecting the ZZ parity check at time $t=7$. 
Figure~\ref{fig:hcf-z-detectors} illustrates a $Z$ measurement error applied to qubits 0 and 2. 
The gray band (see Figure~\ref{fig:hcf-meas}) indicates time step $t=7$ where a parity measurement error occurs at qubits 0 and 2. 
Under time-gapped detector construction (considering Z-type detectors only), detectors compare Z-checks on the same plaquette across non-consecutive time steps, triggering detector D4 (comparing gZZ at $t=7$ with bZZ at $t=11$) and detector D0 (comparing rZZ at $t=3$ with gZZ at $t=7$).

\begin{figure}[H]
\centering
\includegraphics[width=0.85\linewidth]{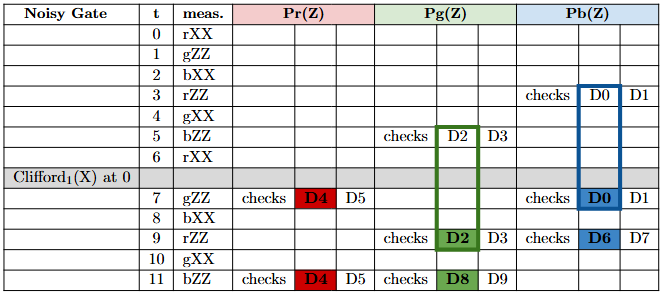}
\caption{Example (c): Data qubit error in HCF code.
The gray band indicates the 
interval $t \in (6,7)$ where a Pauli-X error occurs on qubit 0. 
Colored backgrounds indicate triggered checks (D0 in blue, D2 in green), 
with colored frames showing detector ranges. The X error anti-commutes 
with Z-type checks, creating temporal boundaries that trigger two 
detectors: D0 (spanning rZZ at $t=3$ and gZZ at $t=7$) and D2 
(at $t=9$), yielding $w=2$.}
\label{fig:hcf_data_error}
\end{figure}

\clearpage
%========================
\section{Decoding time}
\label{app:decoding_time}
%========================
As shown in Figure~\ref{fig:decoder_impact_64}, the HCF code shows no discernible difference in logical error rate between MWPM and BP+OSD decoders, whereas the HF code demonstrates that BP+OSD achieves a lower logical error rate than MWPM.

Given that BP+OSD is generally considered computationally more expensive, we evaluated the per-shot decoding time as a function of the physical qubit count $n$ under the EM-ind noise model with $P_{\mathrm{phys}} = 2.2 \times 10^{-3}$.
The decoding time is defined as the wall-clock time from decoder initiation to completion when processing a single pre-generated syndrome serially. This computational result can inform the choice for decoder selection in HF code implementations.

\begin{figure}[H]
  \centering
  \includegraphics[width=8cm]{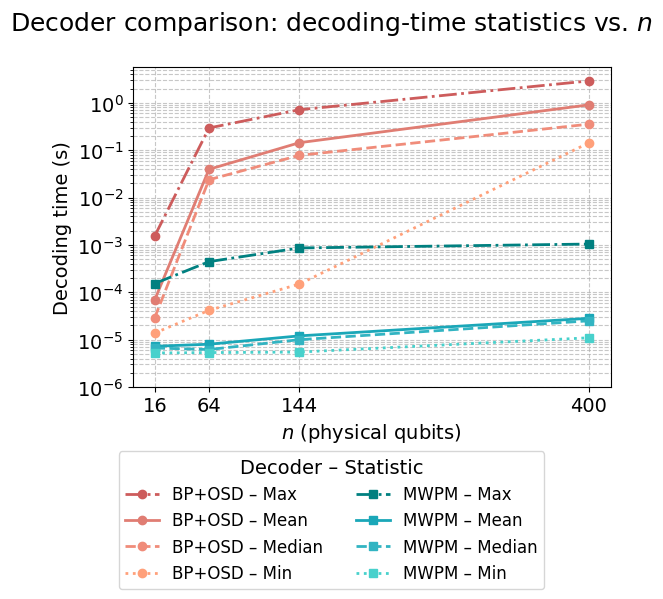}
  \caption{Decoding time comparison between MWPM and BP+OSD decoders. Log-log plot 
  of per-shot decoding time \vs physical qubit count $n$. For each 
  configuration, 10,000 pre-generated syndromes were decoded serially. 
  MWPM requires a mean per-shot time of $28~\mu\mathrm{s}$ at $n=400$, while BP+OSD requires 
  $40~\mathrm{ms}$ at $n=64$ and exceeds $0.9~\mathrm{s}$ at $n=400$.}
  \label{fig:decoding_time_n_plot}
\end{figure}

For each code size and decoder, we sequentially decoded 10,000 syndromes 
and recorded timing statistics (mean, median, minimum, and maximum) for 
each shot. Four code sizes were examined: $n \in \{16, 64, 144, 400\}$, 
corresponding to distances $d \in \{2, 3, 4, 7\}$, respectively. All calculations were performed on an Intel Core i5 processor (2.5 GHz) with 16 GB of RAM. Figure~\ref{fig:decoding_time_n_plot} demonstrates the per-shot decoding time. In the measurements, MWPM remains at $\sim 28~\mu\mathrm{s}$ for $n=400$, whereas BP+OSD requires $\sim 40~\mathrm{ms}$ at $n=64$ and $\sim 0.9~\mathrm{s}$ at $n=400$.

The distribution exhibits a characteristic shape with tightly concentrated min/median/mean values 
and a tail where the maximum values exceed the median by factors of tens to 
hundreds within our simulated parameter range. This distribution reflects the 
algorithmic properties of MWPM: in typical decoding instances, syndrome defects 
form small, localized clusters that require only short-range augmentations and 
a limited number of blossom operations. However, in rare cases where large 
cluster regions emerge (due to locally high defect density or extended 
connectivity patterns), the total number of augmentation and blossom-related 
events increases substantially, resulting in significantly longer computation 
times~\cite{Higgott2025sparseblossom}.

In contrast, BP+OSD exhibits a bimodal distribution structure, with scattered
minimum values (corresponding to cases where BP alone achieves immediate
convergence) and a distinct concentration of higher values when OSD calculation
is required. This bimodal pattern originates from the hybrid architecture of BP+OSD, 
which sequentially combines the computationally efficient BP decoder with the
more computationally intensive OSD procedure when BP fails to converge.

Table~\ref{tab:scaling_fit} presents the scaling analysis of the decoding time, 
which follows the power law $T = \beta n^{\alpha}$. Here, $\alpha$ denotes the 
scaling exponent, $\beta$ represents the prefactor (in microseconds), and $R^2$ 
quantifies the goodness of fit. The mean-based fitting yields the following 
parameters: MWPM ($\alpha = 4.1 \times 10^{-1}$, $\beta = 1.9$) and BP+OSD 
($\alpha = 2.9$, $\beta = 5.2 \times 10^{-2}$).

\begin{table}[H]
  \centering
  \caption{Scaling parameters for decoding time $T = \beta n^{\alpha}$.
  Here $\alpha$ is the scaling exponent, $\beta$ the prefactor in
  microseconds, and $R^2$ the coefficient of determination (goodness of fit for $T$).
  Data from $n = \{16, 64, 144, 400\}$.}
  \label{tab:scaling_fit}
  \footnotesize
  \renewcommand{\arraystretch}{1.1}
  \setlength{\tabcolsep}{6pt}
  \begin{tabular}{l l l l l}
    \hline
    Decoder & Statistic & $\alpha$ & $\beta$ & $R^2$ \\
    \hline
    \multirow{4}{*}{BP+OSD}
  & Max    & 2.3 & $5.6$   & $9.2\times10^{-1}$ \\
  & Mean   & 2.9 & $5.2\times10^{-2}$  & $9.3\times10^{-1}$ \\
  & Median & 2.9 & $2.7\times10^{-2}$  & $9.0\times10^{-1}$ \\
  & Min    & 2.7 & $2.1\times10^{-3}$  & $7.7\times10^{-1}$ \\
\cline{1-5}
\multirow{4}{*}{MWPM}
  & Max    & $6.2\times10^{-1}$ & $3.1\times10^{1}$   & $9.5\times10^{-1}$ \\
  & Mean   & $4.1\times10^{-1}$ & $1.9$   & $8.2\times10^{-1}$ \\
  & Median & $4.0\times10^{-1}$ & $1.7$   & $7.3\times10^{-1}$ \\
  & Min    & $2.1\times10^{-1}$ & $2.6$   & $6.0\times10^{-1}$ \\
\hline
  \end{tabular}
\end{table}

\end{document}